\documentclass{JHEP3}

\pdfoutput=1
\input{epsf.sty}
\usepackage{epsfig,amssymb, graphicx,bm,color}

\def\be{\begin{equation}}
\def\ee{\end{equation}}    
\def\ba{\begin{eqnarray}}
\def\ea{\end{eqnarray}}

\def\2pi{\left(2\pi\right)}

\newcommand{\Expect}[1]{\left\langle #1 \right\rangle}

\renewcommand{\k}{\ensuremath{\bm{k}}}
\newcommand{\fnl}{\ensuremath{f_{\rm NL}}}
\newcommand{\hinvmpc}{\,h^{-1}{\rm Mpc}}
\newcommand{\hmpcinv}{\,h{\rm \,Mpc^{-1}}}
\newcommand{\Msunhinv}{\,h^{-1}M_\odot}
\newcommand{\Msun}{\,M_\odot}

\title{A generalized local ansatz and its effect on halo bias} \author{Sarah
  Shandera${}^{1}$, Neal Dalal${}^{2}$ and Dragan Huterer${}^{3}$
  \\
$^1$ Perimeter Institute for Theoretical Physics, Waterloo, Ontario, Canada;\\
$^2$ Canadian Institute for Theoretical Astrophysics, University of Toronto, Toronto, Ontario, Canada;\\
$^3$ Department of Physics, University of Michigan, Ann Arbor, Michigan, USA\\ 

  }
  
  \abstract {Motivated by the properties of early universe scenarios that
    produce observationally large local non-Gaussianity, we perform N-body
    simulations with non-Gaussian initial conditions from a generalized local
    ansatz. The bispectra are schematically of the local shape, but with
    scale-dependent amplitude. We find that in such cases the size of the
    non-Gaussian correction to the bias of small and large mass objects
    depends on the amplitude of non-Gaussianity roughly on the scale of the
    object. In addition, some forms of the generalized bispectrum alter the
    scale dependence of the non-Gaussian term in the bias by a fractional
    power of $k$. These features may allow significant observational
    constraints on the particle physics origin of any observed local
    non-Gaussianity, distinguishing between scenarios where a single field or
    multiple fields contribute to the curvature fluctuations. While analytic predictions for the
    non-Gaussian bias agree qualitatively with the simulations, we find
    numerically a stronger observational signal than expected. This suggests
    that a more precise understanding of halo formation is needed to fully
    explain the consequences of primordial non-Gaussianity.}

\preprint{}


\begin{document}
\maketitle

\section{Motivation}
Non-Gaussianity that originates from the inflationary epoch leaves distinct
signatures in present-day astrophysical measurements, and therefore
provides a unique link to the early universe. Interactions of the field(s)
sourcing the primordial curvature fluctuations introduce non-Gaussian imprints
in the statistics of the temperature fluctuations in the Cosmic Microwave
Background (CMB) and of the density fluctuations that collapse into bound
objects. These effects give us many independent probes of the signals of
inflationary physics at different redshifts, sensitive to a range of
scales. While current measurements from the CMB confirm that the spectrum of
primordial fluctuations is Gaussian to a remarkable part in $10^3$, that bound
is still four orders of magnitude away from testing primordial non-Gaussianity
at the level predicted by slow-roll inflation and more than one order of
magnitude above the level expected from non-linear post-inflationary
processing of the fluctuations (see eg \cite{Pitrou:2010sn} for a recent
calculation). In addition, {\it any} deviation from the simplest single field
slow-roll inflationary scenario, including multiple fields, derivative
interactions, features in the potential, or non-Bunch-Davies initial
conditions (see \cite{Komatsu:2009kd} for a summary) can lead to observable
non-Gaussianity at levels within current constraints but well above the
slow-roll prediction. Upcoming data from the Planck satellite \cite{Planck}
and a variety of galaxy surveys
\cite{Carlstrom:2009um,BOSS,DES,HETDEX,PAU,LSST} have the potential to achieve accuracy on non-Gaussianity at the level expected from non-linear evolution alone. For recent reviews, see
\cite{Chen_AA,Bartolo_AA,Liguori_AA,Verde_AA, Desjacques:2010jw}.

Primordial non-Gaussianity is most effectively constrained by complementary
measurements from the CMB and Large Scale Structure (LSS). The galaxy power
spectrum and bispectrum and cluster number counts provide independent
statistics with different systematics, sensitive to different qualitative
features of the primordial non-Gausianity. Combined with the CMB, these observations will constrain a wide range of
qualitative features of any observed non-Gaussianity (including amplitude,
shape, sign and scale dependence) which can rule out large classes of
inflationary models.

Non-Gaussianity of the local type (with bispectrum maximum in the squeezed
limit; $k_1\approx k_2\gg k_3$) has recently generated a good deal of interest
in part because it will be especially well-constrained by LSS observations
\cite{Dalal:2007cu}. Even in the case of Gaussian fluctuations, the statistics
of collapsed objects are different from those of the underlying density field,
and the ratio of the clustering of the two is known as the halo bias
\cite{Kaiser:1984sw}. The particular coupling of long and short wavelength
modes in local non-Gaussianity introduces an additional, distinctive
correction (proportional to $1/k^2$) in the power spectrum of collapsed
objects which will allow strong observational constraints on local
non-Gaussianity \cite{Dalal:2007cu,Slosar:2008hx}. From a theoretical point of
view, observably large primordial non-Gaussianity of this type {\it requires}
at least two fields to contribute to the scenario - single field inflation
alone can only generate a bispectrum of the local shape with an extremely
small amplitude (of order the spectral index of the primordial power spectrum
\cite{Creminelli:2004yq}). The complete phenomenology of multi-field models is
rich but we will show here that there {\it are} qualitative differences that
are observationally distinguishable in the halo bias. We propose here a generalization of the local ansatz that is phenomenologically useful and captures the physics of many possible multi-field models. The generalized ansatz allows for different types of scale-dependent amplitude $\fnl$ along with the standard local shape.

While signatures of primordial non-Gaussianity in LSS can often be predicted
analytically, accurate comparisons of observables with theoretical predictions
require the intermediate step of numerical simulations to validate or correct
any the analytical relations. In this paper we build on previous work of Dalal
et al.\ \cite{Dalal:2007cu} to numerically investigate the effect of
scale-dependent, local non-Gaussian initial conditions. Interestingly, we find
theoretically and numerically that the halo bias is sensitive to two different
types of scale dependence that can constrain and distinguish between
inflationary models. However unlike in the constant $\fnl$ case, the simplest
theoretical prediction for the bias in models with scale-dependent
non-Gaussianity does not fully agree with our numerical results. In this paper we will motivate our new
non-Gaussian ansatz, present the analytic predictions from that model and the
associated simulations. We will discuss a possible explanation for the
discrepancy, which indicates that this problem constitutes an interesting test for our understanding of structure growth, although we postpone a detailed
analysis for future work.

The paper is organized as follows. In Section \ref{sec:model} we discuss in
more detail the motivation from inflationary theory. A self-contained and
purely phenomenological discussion starts in Section \ref{sec:LSS}, where we
use the peak-background split method to demonstrate the qualitative ways
scale-dependent non-Gaussianity may be observable. We also present forecasts
for differentiating the bispectra based on the analytic predictions. In
Section \ref{sec:sim} we show the results of numerical simulations, which
demonstrate a stronger signal than the analytic prediction, and so are encouraging for the observational prospects. We speculate on a possible explanation for the discrepancy between theory and simulation and then conclude in Section \ref{sec:concl}.

\section{A bigger family for the local ansatz}\label{sec:model}

The original ``local ansatz'' to add non-Gaussianity to the primordial
perturbations is \cite{Salopek:1990jq, Verde:1999ij, Komatsu:2001rj}: 
\begin{equation}
\Phi({\bf x})=\Phi_G({\bf x})+\fnl\left[\Phi_G^2({\bf  x})-\langle\Phi_G^2({\bf x}) \rangle\right] +\dots,
\label{eq:NG1}
\end{equation} 
where $\Phi({\bf x})$ is (minus) the gravitational potential, $\Phi_G({\bf
  x})$ is a Gaussian random field and the degree of non-Gaussianity is
parameterized by (typically constant) $\fnl$. Here a positive $\fnl$ leads to
a positive skewness in the density perturbations, and so more very large
objects, in the same sign convention as WMAP \cite{Komatsu:2003fd}\footnote{We
  use the convention that $\fnl$ is defined in terms of (minus) the
  gravitational potential early in the matter era. We caution that there is another convention (used, for example, in
  \cite{Grossi:2007ry, Carbone:2008iz}) that defines $\fnl^{\rm LSS}$ in terms
  of the gravitational potential normalized to present day amplitude, which is
  related to the WMAP convention used here by $\fnl^{\rm LSS}=\fnl
  (g(z=\infty)/g(z=0))$ ($\approx 1.36\fnl$ in the WMAP7 cosmology), where $g(z=0)/g(\infty)$
    is often referred to as the growth suppression factor.}.

In many scenarios the primary effect of the non-Gaussian correction appears as a non-zero bispectrum, defined as
\begin{eqnarray}
\langle\Phi(\k_1)\Phi(\k_2)\Phi(\k_3)\rangle&\equiv&(2\pi)^3\,\delta_D^3(\k_1+\k_2+\k_3)B_{\Phi}(\k_1,\k_2,\k_3)\;.
\end{eqnarray}
For the local ansatz above the bispectrum is
\begin{eqnarray}  
B_{\Phi}(\k_1,\k_2,\k_3)&=&\fnl\left[2\:P_{\Phi}(\k_1)P_{\Phi}(\k_2)+2\:\textrm{perm.}\right ],  
\end{eqnarray} 
where as usual we define
\begin{equation}
  \label{eq:Pzetaofk}
\Expect{\Phi(\k_1)\Phi(\k_2)}\equiv(2\pi)^3\,\delta_D^3(\k_1+\k_2)P_{\Phi}(k_1)=(2\pi)^3\,\delta_D^3(\k_1+\k_2)\frac{2\pi^2\Delta_{\Phi}^2(k_1)}{k_1^3}.
\end{equation} 
The subscript $D$ distinguishes the Dirac delta function from the density
perturbation. Assuming the spectral index, $n_s$, has no significant $k$-dependence,
the dimensionless power spectrum is given by $\Delta_{\Phi}^2=A_0(k/k_0)^{n_s-1}$.

CMB data (WMAP7) already constrain $-10<\fnl< 74$ at 95\% confidence
(\cite{Komatsu:2010fb}; see also \cite{Creminelli:2006rz,Smith:2009jr}) and could
potentially achieve $\Delta \fnl\sim $ few from the Planck satellite
\cite{Komatsu:2003fd, Creminelli:2005hu, Spergel:2006hy, Chen:2006ew}. The
best current constraint from LSS comes from the scale-dependent bias induced
in the galaxy power spectrum, giving $-29<\fnl< 69$ at 95\% CL
(\cite{Slosar:2008hx}; see also \cite{Afshordi:2008ru}).

While the local ansatz is a useful phenomenological tool, it is only a first
step toward modeling and constraining primordial non-Gaussianity motivated by
the fundamental physics of inflation. Specifically, the local ansatz resembles
the first term in a series that arises from the transfer of isocurvature to
curvature fluctuations during or at the end of inflation. Such a transfer may
be due to additional scalar fields during inflation (multi-field
\cite{Rigopoulos:2005us, Seery:2005gb, Vernizzi:2006ve, Battefeld:2006sz,
  Choi:2007su, Yokoyama:2007uu, Yokoyama:2007dw, Sasaki:2008uc, Cogollo:2008bi, Rodriguez:2008hy, Byrnes:2008wi,
  Byrnes:2008zy, Byrnes:2009pe, Byrnes_AA,Huang:2010es}), or after (the curvaton scenario
\cite{Lyth:2002my, Ichikawa:2008iq, :2008ei, Byrnes:2009pe, Huang:2008zj, Enqvist:2009ww,
  Enqvist:2009eq, Alabidi:2010ba, Chambers:2009ki}), or inhomogeneous reheating \cite{Dvali:2003em,Zaldarriaga:2003my}. 
  
 There are at least four possible sources of non-Gaussianity that generate
bispectra that are largely well-captured by the local ansatz shape in the final curvature
perturbations. First, a spectator field during inflation is not constrained to
have a flat potential, so there may be intrinsic non-Gaussianity in that field
that is not tightly constrained by the slow-roll conditions and which can be
transferred to non-Gaussianity in the curvature. Second, in multi-field models non-linear evolution
of curvature modes outside the horizon will generate non-Gaussianity in the
observed curvature perturbations even if the field(s) themselves have no
interactions other than gravitational \cite{Byrnes:2009pe}. Third, the
conversion of curvaton isocurvature fluctuations to curvature after inflation
depends on the energy density in the curvaton field, which is at least
quadratic in the fluctuations and so introduces non-Gaussianity of the local
type \cite{Lyth:2002my}. Finally, loop corrections may, in special cases,
generate a scale-dependent non-Gaussianity \cite{Kumar:2009ge}.

Phenomenologically, we can write a more general ansatz for the bispectrum of (minus) the gravitational potential that is factorizable and symmetric in momentum by introducing two functions, $\xi_s(k)$ and $\xi_m(k)$:
\begin{equation}
\label{eq:runfNLpheno}
B_{\Phi}(\k_1,\k_2,\k_3)=\xi_{s}(k_3)\xi_{m}(k_1)\xi_{m}(k_2)P_{\Phi}(k_1)P_{\Phi}(k_2)+5\:{\rm
  perm}\;.  
\end{equation} 
This ansatz captures a wide range of physically motivated and
perturbatively controlled models, where the functions $\xi_{s,m}$ are at most
weak functions of scale. The notation refers to the physical origin of the two
functions in inflationary scenarios: $\xi_s$, with {\it s} for single field, is
different from one if one of the fields has non-trivial self interactions or nonlinearly sources curvature perturbations;
$\xi_m$, with {\it m} for multi-field, is different from one when two or more
fields both contribute to the power in curvature fluctuations. We will discuss
several illustrative examples next.

\subsection{Two field inflation}
First, we consider a two field inflation scenario where running
non-Gaussianity can be obtained, following \cite{Vernizzi:2006ve,
  Byrnes:2008zy}. In the $\delta N$ formalism \cite{Sasaki:1995aw,
  Lyth:2005fi}, one uses the dependence on the number of e-folds of inflation
on the fields present to relate the curvature fluctuations to the scalar field
fluctuations. Even if a field does not source the inflationary Hubble
parameter $H$, the point where inflation ends (and so the number of e-folds,
$N$) can still depend on the position of the field. Then, we can express the
curvature perturbation resulting from fluctuations of two fields $\phi$ and
$\sigma$ (up to second order) as
\begin{equation}
\zeta(k)=N_{,\phi}(k)\delta\phi(k)+N_{,\sigma}(k)\delta\sigma(k)+
\frac{1}{2}N_{,\sigma\sigma}(k)[\delta\sigma\star\delta\sigma](k)\;
+\dots\
\end{equation}
where for simplicity we have assumed one of the non-Gaussian terms ($N_{,\sigma\sigma}$)
dominates the other ($N_{,\phi\phi}$) and $N_{,\phi\sigma}=0$. All quantities are evaluated at
horizon crossing for the mode $k$, and $N_{,\phi}$ is the derivative of the
number of e-folds with respect to the field $\phi$. To gain some intuition
about the pattern of multiplications and convolutions in this expression, recall that in single field inflation the running of
$N_{,\phi}=-\frac{H}{\dot{\phi}}\sim1/\sqrt{\epsilon}$ contributes the term
proportional to $\eta$ to the spectral index. In addition, the form of the quadratic term generates the standard result that 
the bispectrum in the squeezed limit goes like the spectral index, $n_s-1$,
evaluated at horizon crossing of the short wavelength (large $k$) modes
\cite{Maldacena:2002vr} (although one must take into account pre-horizon crossing non-Gaussianity generated in the statistics of the field $\delta\phi$ to get the complete bispectrum correct). In the single field case, the amplitude of the non-Gaussianity in the
curvature perturbation is small, but it does run if the spectral index runs and
the dominant term has scale-dependent $\fnl$ evaluated at the scale of the
short modes\footnote{A detailed discussion of how this is consistent with real-space formulations of the local ansatz can be found in \cite{Byrnes:2010ft}.}.

In a multi-field scenario, the modes for each individual field have fluctuations of order $H$, so that
\begin{eqnarray}
\Expect{\delta\phi({\bf k})\delta\phi({\bf k}^{\prime})}=
\Expect{\delta\sigma({\bf k})\delta\sigma({\bf k}^{\prime})}
&=&(2\pi)^3\delta_D^3({\bf k}+{\bf k}^{\prime})\frac{(2\pi^2)}{k^3}\frac{H_{*}^2}{4\pi^2}\\[0.2cm]\nonumber
&\equiv&(2\pi)^3\delta_D^3({\bf k}+{\bf k}^{\prime})P(k)\;.
\end{eqnarray}
where the asterisk is a reminder that $H$ is evaluated at horizon-crossing for each wavenumber $k$. Then the total curvature power spectrum can be written
\begin{eqnarray}
\Expect{\zeta({\bf k})\zeta({\bf k}^{\prime})}&\equiv&(2\pi)^3\delta_D^3({\bf k}+{\bf k}^{\prime})P_{\zeta}(k)\\[0.2cm]\nonumber
&=&(2\pi)^3\delta_D^3({\bf k}+{\bf k}^{\prime})P(k)(N_{,\phi}^2+N_{,\sigma}^2)\\[0.2cm]\nonumber
&=&(2\pi)^3\delta_D^3({\bf k}+{\bf k}^{\prime})[P_{\zeta(\phi)}+P_{\zeta(\sigma)}]\;.
\end{eqnarray}
The tree-level bispectrum, assuming $\langle\delta\phi\delta\sigma\rangle=0$, is
\ba
\label{eq:SymFact}
B_{\zeta}({\bf k}_1,{\bf k}_2,{\bf k}_3)&=&\frac{1}{2}
N_{,\sigma\sigma}(k_3)\:\frac{P_{\zeta(\sigma)}(k_1)}{P_{\zeta}(k_1)}\frac{P_{\zeta(\sigma)}(k_2)}{P_{\zeta}(k_2)}\:P_{\zeta}(k_1)P_{\zeta}(k_2)+5\:\textrm{perm.},\\\nonumber
B_{\Phi}({\bf k}_1,{\bf k}_2,{\bf k}_3)&\equiv&
\xi_s(k_3)\xi_m(k_1)\xi_m(k_2)\:P_{\Phi}(k_1)P_{\Phi}(k_2)+5\:\textrm{perm.},
\ea
where $N_{,\sigma\sigma}$ depends on non-trivial self (and gravitational) interaction terms of just the field $\sigma$, so we relabel it $\xi_s$, with $s$ for single field. The fraction of power in the $\sigma$ field is different from one  only if both fields contribute significantly to the power in fluctuations so we have labeled this function with an $m$ for multi-field. (Otherwise, the bispectrum would reduce to the usual single-field expression, where $\fnl$ must be of order slow-roll - that is, the term quadratic in $\delta\phi$ would be most important, giving a bispectrum with the same form as the first line of Eq.(2.9) but with the coefficient of the power spectrum terms $N_{\phi\phi}(k)$.) Assuming that the potential $\Phi$ is defined in the matter era, the precise relationship between the first and second lines above is
\ba
\frac{5}{6}N_{,\sigma\sigma}(k)&=&\xi_s(k)\\\nonumber
\frac{P_{\zeta(\sigma)}(k)}{P_{\zeta}(k)}&=&\xi_m(k)\;.
\ea

We note that quite generally all the coefficients $N_{,\phi}$, $N_{,\sigma}$,
$N_{,\sigma\sigma}$, etc will be scale-dependent as the potentials for the
fields are not exactly flat. In that sense, in any two-field
scenario with large local non-Gaussianity, running of the amplitude through
the function $\xi_s(k)$ is as natural as running of the spectral index. It may
be somewhat fine-tuned to have two fields contribute to the amplitude of
fluctuations (although this is hard to say in the absence of compelling
particle physics realizations of inflation), but if they do it is likely
natural for their potentials to be slightly different so that $\xi_m(k_2)$ is
scale-dependent. We will parametrize this scale dependence by writing
\begin{equation}
\xi_{s,m}(k)=\xi_{s,m}(k_p)\left(\frac{k}{k_p}\right)^{n^{(s),(m)}_f}
\label{eq:xi_powerlaw}
\end{equation}
where $k_p$ is a (theoretically irrelevant) pivot point.

\subsection{Mixed curvaton/inflaton scenario}
Now suppose the curvature perturbation comes partly from a Gaussian inflaton
field ($\phi$) and partly from a `curvaton' field ($\sigma$) which was a
spectator during inflation but contributes to the curvature perturbation
afterwards \cite{Mollerach,Linde:1996gt,Moroi:2001ct, Enqvist:2001zp,
  Lyth:2001nq, Lyth:2002my}. The curvaton naturally has a contribution that is
quadratic in real space since it contributes proportionally to the energy
density in its fluctuations. Assuming a purely quadratic potential for
the curvaton gives
\begin{eqnarray}
\rho_{\sigma}=\frac{1}{2}m_{\sigma}^2(\sigma+\delta\sigma)^2\Rightarrow\delta\rho_{\sigma}=\frac{1}{2}m^2(2\sigma\delta\sigma+\delta\sigma^2).
\end{eqnarray}
The field fluctuations are still generated during inflation, with amplitude
$\sqrt{\langle\delta\sigma^2\rangle}=H/2\pi$. The quadratic term means that
the curvaton can contribute a local-type non-Gaussianity with $\fnl$
constant and determined by the proportion of energy in the curvaton at the
time the field decays.

Then we can write the total curvature field as a sum of contributions from the inflaton and curvaton:
\begin{equation}
\zeta(x)=\zeta_{\phi}(x)+\zeta_{\sigma}(x)+
\frac{3}{5}\fnl^{\sigma}(\zeta_{\sigma}(x)^2-\langle\zeta_{\sigma}(x)^2\rangle)
\end{equation}
where the factor of $3/5$ enters since $f_{NL}$ is conventionally defined for the matter era potential (Eq.(\ref{eq:NG1})) rather than the primordial curvature.

Assuming the fields don't couple, the bispectrum takes the familiar local form,
but now in terms of $P_{\zeta(\sigma)}$:
\begin{equation}
B_{\zeta}(\k_1,\k_2,\k_3)=\frac{3}{5}\fnl^{\sigma}[P_{\zeta(\sigma)}(k_1)P_{\zeta(\sigma)}(k_2)+5\:{\rm
    sym}]\;.
\end{equation}
If we define the ratio of power contributed by the curvaton
\begin{equation}
\xi(k)=\frac{P_{\zeta(\sigma)}(k)}{P_{\zeta(\sigma)}(k)+P_{\zeta(\phi)}(k)}=\frac{P_{\zeta(\sigma)}(k)}{P_{\zeta}(k)}
\end{equation}
we can write
\begin{eqnarray}
\label{eq:runfNLcurv}
B_{\zeta}(\k_1,\k_2,\k_3)&=&\frac{3}{5}\fnl^{\sigma}[\xi(k_1)\xi(k_2)P_{\zeta}(k_1)P_{\zeta}(k_2)+5\:{\rm perm}]\\\nonumber
B_{\Phi}(\k_1,\k_2,\k_3)&\equiv&\xi_m(k_1)\xi_m(k_2)P_{\Phi}(k_1)P_{\Phi}(k_2)+5\:{\rm perm}
\end{eqnarray}
where we have absorbed $f_{NL}^{\sigma}$ into $\xi_m$, that is
$\xi_m(k)=\sqrt{f_{NL}^{\sigma}}\xi(k)$. Again, we parametrize the function
$\xi_m(k)$ as a simple power law, $\xi_m(k)\propto k^{n_f^{(m)}}$.

\subsection{Curvaton alone}
If non-Gaussianity comes from the curvaton alone, and a potential other than
quadratic is considered, the bispectrum can again take the form
\cite{Byrnes:2010xd,Huang:2010cy} 
\ba
\label{eq:CurvOnly}
B_{\Phi}({\bf k}_1,{\bf k}_2,{\bf k}_3)&\equiv&
\xi_s(k_3)P_{\Phi}(k_1)P_{\Phi}(k_2)+5\:\textrm{perm.},
\ea
where $\xi_s(k)$ can be parametrized as a power law, at least for some potentials, and $n_f^{(s)}$ apparently can have either sign. Inhomogeneous reheating may similarly generate this bispectrum \cite{Byrnes:2010ft}.

\subsection{Relation to the spectral index}
The running of $\xi_m(k)$ is an important physical feature of either type of
two-field model: it is the evolution of the relative power in the two fields
during inflation. Just as the spectral index measures the variation of the
overall amplitude of fluctuations during inflation, for two-field models the
bispectral index $n_f^{(m)}$ can provide complementary information about how the
contribution from each field evolves. For some curvaton scenarios there would
be a link between running non-Gaussianity and large scale power asymmetry in
the CMB \cite{Erickcek:2009at}.

There is a precise relationship between the spectral index and the bispectral index $n_f^{(m)}$:
\begin{eqnarray}
\frac{d\ln P_{\zeta}}{d\ln k}&\equiv&n_s-1\\[0.2cm]\nonumber
\frac{d\ln P_{\zeta(\sigma)}}{d\ln k}&\equiv&n_{\sigma}-1\\[0.2cm]\nonumber
\frac{d\ln \xi_m}{d\ln k}&\equiv&n_f^{(m)}=n_{\sigma}-n_s
\end{eqnarray}
Although the running of the bispectrum may have either sign, models with a red tilt for the field $\sigma$ are anecdotally more common and in that case we have
\begin{equation}
n_f^{(m)}\leq-(n_s-1)\;.
\end{equation}
Notice that some of the literature (e.g. \cite{Byrnes:2009pe}) defines
$\fnl(k)=\xi_m(k)^2$ and so quotes $n_f\leq-2(n_s-1)$. Here
however, we will see that there are two different shifts in the non-Gaussian
bias, each dependent on one factor of $\xi_m(k)$, so we define $n_f^{(m)}$ as the
running in that function. Finally, notice that the spectral index of the
observed curvature perturbation depends on the running of both fields $\phi$
and $\sigma$. If the running of the fields is large enough, it will change
which field dominates the curvature statistics. 

Whatever the origin of the running in either function $\xi_{m,s}$, it parametrizes the deviation from exactly quadratic potentials in either field and so is expected to be generically on the order of slow-roll parameters (and should be
to avoid substantial corrections to this parameterization). We will use
somewhat large values of the running to confirm the behavior of this type of
model in our simulations, but the observational goal should be to measure
$|n_f^{(s),(m)}|\sim\mathcal{O}(n_s-1)$. We discuss the potential of future
surveys to reach this goal in Section 5. For the standard quadratic curvaton case, $n_f^{m}>0$
seems more natural (that is, non-Gaussianity increases on small scales) while
Byrnes et al.\ \cite{Byrnes:2008zy} found $0>n_f^{(s)}\gtrsim-0.1$ in a survey
of multi-field hybrid inflation models. The sign can be understood if the
non-Gaussianity is entirely due to non-linear evolution outside the
horizon. Then large scale modes (which exit earlier) will to be more
non-Gaussian than smal scale modes.\footnote{Previous authors have employed
  different notation for scale-dependent local models. In particular, Byrnes
  et al, in an extensive discussion of possible multi-field bispectra
  \cite{Byrnes:2010ft} propose a definition of $\fnl$ and its running that for
  our ansatz correspond to $f^{\rm
    Byrnes}_{NL}(k_1,k_2,k_3)\equiv [\xi_s(k_3)\xi_m(k_1)\xi_m(k_2)P_{\Phi}(k_1)P_{\Phi}(k_2)+{\rm
      sym}] / [P_{\Phi}(k_1)P_{\Phi}(k_2)+{\rm sym}]$ and $n^{\rm
    Byrnes}_{\fnl}\equiv d\ln|\fnl^{\rm Byrnes}(k_1=k_2=k_3=k)|\,/\,d\ln
    k=n_f^{(s)}+2n_f^{(m)}$.}
\footnote{ In a discussion of the ability of observations to constrain
  two-field models of the mixed curvaton/inflaton type, Tseliakhovich et al
  \cite{Tseliakhovich:2010kf} recently defined a variable $x_1$ where
  $x_1=\xi_m^2$ and where only the scale-independent case was considered. In
  addition, their function $\xi$ is defined differently: $\xi^{\rm
    here}=1/(1+(\xi^{\rm there})^2)$.}

\subsection{A comment on naturalness and completeness}

Given that it is already difficult to convincingly explain one field with a
very flat potential, we may reasonably ask if the scenarios we are considering
are even less likely than the usual single-field inflation. It is very hard to
answer that question without better fundamental models - it may be that where
there is one inflaton-like field, there are naturally several (especially in
higher dimensional models), or not. In inflation, there is a very compelling
reason why the spectral index should be slightly different from one: old
inflation models with exact de Sitter space are difficult to connect to the
early, hot universe after inflation, while slow-roll with the Hubble parameter
not exactly constant can have a natural end to inflation and a period of
reheating. {\it If} the slow-roll scenario is right and {\it if} two fields
are present and relevant during inflation, it may be reasonable to expect that
they both have nearly flat and yet not identical potentials. If one accepts
that local type non-Gaussianity is natural (or more compellingly, if it is
observed), scale dependence is also natural. In the absence of a range of
compelling high energy models, it is hard to quantify the likelihood of any of
these scenarios.

However, from a phenomenological point of view, considering a generalized
local ansatz is helpful in two ways: first, it argues for a careful analysis
of different mass tracers in any test for primordial local type
non-Gaussianity and second, it provides a test of our understanding of
structure formation. As we will see, the existing expressions for halo bias do
not give particularly satisfactory agreement with our simulations.

The generalized local ansatz above is useful to uncover new observational
signatures, and it would be interesting to investigate to what extent it holds
in more complicated models with more (and coupled) fields. However, even with this ansatz we
are still far from considering all possible known effects. In most two field models,
we expect higher order terms (like a $\zeta^3$ contribution) to be
present in the expression for the non-Gaussian curvature. Those corrections
are also important for comparison of observation with realistic models and
have been considered in \cite{Seery:2006js, Ichikawa:2008ne, Jeong:2009vd,
  Desjacques:2009jb, Chingangbam:2009vi,Enqvist:2009ww}. In addition, there
are other possibilities that require something even more general than the
symmetric, factorizable form. For example, non-gaussianity generated by loop
effects can sometimes be large and goes like \cite{Kumar:2009ge}
\begin{equation}
\langle\zeta^3\rangle\propto \fnl({\rm min}\{k_1,k_2,k_3\})[P(k_1)P(k_2)+5\:{\rm perm}.]
\end{equation}
In addition, the power-law behavior of our ansatz is a poor model for scenarios with a feature at some particular scale, such as \cite{Chen:2008wn} (features in the potential) or \cite{Riotto:2010nh} where
non-Gaussianity effectively switches on at some scale where a spectator
field becomes light.

Finally, we note that scale-dependent non-Gaussianity may also arise in
other ways and for other bispectra, but most other examples in standard inflationary models are less divergent in the squeezed limit than the local shape is and so have a weaker signal in the bias. However, there is a small region of parameter space in ekpyrotic models that seems to generate bispectra with scale-dependent amplitudes consistent with current observations, and {\it more} divergent than the local ansatz \cite{Khoury:2008wj}. 

\section{Generalized local ansatz and large scale structure statistics}\label{sec:LSS}

Our ansatz for the factorizable, symmetric, scale-dependent local bispectrum is
\begin{equation}
B_{\Phi}(\k_1,\k_2,\k_3)=\xi_{s}(k_3)\xi_{m}(k_1)\xi_{m}(k_2)P_{\Phi}(k_1)P_{\Phi}(k_2)+5\:{\rm
    perm}\;.
\end{equation}
where we parametrize the $k$-dependence of the amplitude as
\begin{equation}
\label{eq:runfNL}
\xi_{s,m}(k)=\xi_{s,m}(k_p)\left(\frac{k}{k_p}\right)^{n^{(s),(m)}_f}\;.
\end{equation}
with $|n_f^{(s),(m)}|<1$. Ideally, the pivot scale $k_p$ can be chosen at a
point where the amplitude and running of the shape are as close to
uncorrelated as possible. We adopt $k_p=0.04\: {\rm Mpc}^{-1}$ based on
analysis for the CMB in \cite{Sefusatti:2009xu}, although they used a slightly
different ansatz for the scale dependence. Note, however, that
the constraints on $\xi_{s,m}(k)$ will be entirely independent of the chosen
value of $k_p$.

Although generically we might expect both functions $\xi_s(k)$ and $\xi_m(k)$
to be present, we can consider the two functions separately for
simplicity. For that reason, we will compare the following two bispectra in
what follows:
\begin{eqnarray}
\label{eq:twomodels}
B^{s}_{\Phi}(\k_1,\k_2,\k_3)&=&\xi_s(k_1)P_{\Phi}(k_2)P_{\Phi}(k_3)+5\:{\rm perm}
\\[0.2cm]\nonumber
B^{m}_{\Phi}(\k_1,\k_2,\k_3)
&=&\xi_m(k_1)\xi_m(k_2)P_{\Phi}(k_1)P_{\Phi}(k_2)+5\:{\rm  perm}
\end{eqnarray}
The first line applies to a model where only one field contributes to the curvature perturbations (and the inflationary background is sourced by something else). For example, it is generated by a curvaton model where the potential has
terms other than the mass term, Eq.~(\ref{eq:CurvOnly}), or from a simplified version of the $\delta N$
case from Section 2 (where we take $\xi(k)$, the ratio of power in the two
fields, to be constant). Since the curvature perturbations come only from one field, we label the new function $\xi_s(k)$ with $s$ for single field. Scale-dependence in this function indicates the presence of non-trivial self-interactions (eg, deviation of the curvaton potential from exactly quadratic). The second line corresponds to a scenario where (at least) two fields contribute to the curvature perturbations, but the relevant self-interactions are purely quadratic. For example, this is the mixed inflaton/curvaton model of Eq.~(\ref{eq:runfNLcurv}), where the curvaton has only a quadratic potential. The label $m$ on $\xi_m(k)$ indicates that multiple fields contribute to the curvature perturbations. Scale-dependence in $\xi_m$ shows how much the potentials for the fields differ. 

Notice that the first model in Eq.(\ref{eq:twomodels}) has a form that is
equivalent to the bispectrum one would get from generalizing the local ansatz
by
\begin{eqnarray}
\label{eq:runfNLDeltaN}
\Phi({\bf x})&=&\Phi_G({\bf x})+\fnl\ast\left[\Phi_G^2({\bf x})-\langle\Phi_G^2({\bf x}) \rangle\right] \; .\
\end{eqnarray}
This also justifies the single-field label. Scale dependence of this type was studied recently in Ref.~\cite{Becker}.

\subsection{Scale-independent non-Gaussianity and bias}\label{subsec:bias}

In this section we will work out a prediction for the possible
signatures of our generalized local ansatz in the halo power
spectrum. We will only be concerned with the behavior of the power
spectrum at very small $k$, where the deviation from the Gaussian case
is largest. The matter perturbations $\delta$ at redshift $z$ are related to the perturbations in the early matter era potential $\Phi$ by
\begin{eqnarray}
\delta(\vec{k},z)&=&M(k,z)\Phi(\vec{k})\\[0.2cm]\nonumber
M(k,z)&=&\frac{2}{3}\frac{1}{\Omega_m}\frac{c^2}{H_0^2}D(z)\frac{g(0)}{g(\infty)}T(k)k^2,
\end{eqnarray}
so that $P_{\delta}(k,z)=M^2(k,z)P_{\Phi}(k)$. Here $\Omega_M$ is the matter
density relative to critical, $H_0$ is the Hubble constant, $D(z)$ is the
linear growth function at redshift $z$ normalized to one today, and the
growth suppression factor is $\frac{g(z=0)}{g(z=\infty)}\simeq0.76$ in the
best-fit $\Lambda$CDM model. We use the Eisenstein \& Hu \cite{Eisenstein:1997jh} fit to
the transfer function $T(k)$. The variance of density fluctuations at 
redshift $z$ smoothed on a scale $R$ associated to mass $M$ is 
$\sigma^2(M,z)$, defined by
\begin{equation}
\sigma^2(M, z)=\int_0^{\infty}\frac{dk}{k}W_R(k)^2M(k,z)^2\Delta^2_{\Phi}(k).
\end{equation}
where the power spectrum of the primordial curvature perturbations is given by
Eq.~(\ref{eq:Pzetaofk}) and $W_R(k)$ is the Fourier transform of the top-hat
window function.  The spatial smoothing scale $R$ is related to the smoothing
mass scale $M$ via
\begin{equation}
M = {4 \over 3} \pi R^3 \rho_{m, 0},
\label{eq:mass_scale}
\end{equation}
where $\rho_{m, 0}$ is the matter energy density today. We write the combination $M(k,z)W_R(k)\equiv M_R(k,z)$.

\subsection{Peak-background split and halo bias}\label{subsec:peak_back}

Halos in N-body simulations are associated with peaks of the 
initial, linear density field $\delta\propto k^2\Phi$,
whose heights exceed some threshold 
\cite{Dalal:2008zd, Robertson:2008jr}.  The basic idea of the
peak-background split \cite{Cole:1989vx} is to compute the effect
of long-wavelength background modes on the heights of small-scale
peaks, and thereby estimate the large-scale clustering of halos.  The
procedure used in the peak-background split is to perturb a single
background mode $\Delta\Phi(\k_l)$ and propagate the effect of this
perturbation to the height of a peak near threshold.   In
Gaussian cosmologies, where there is no mode coupling, the heights of peaks
are simply boosted by the density associated with the background mode,
$\Delta\delta(\k_l)\propto k_l^2 \Delta\Phi(\k_l)$.  
With non-Gaussianity, however,
there is mode coupling, so we have to compute how this background mode affects
shorter-wavelength modes $\Phi(\k_s)$ as well.  This clearly involves looking
at the bispectrum in the {\em squeezed} limit
$B_{\Phi}(\k_l,\k_s,-\k_s-\k_l\approx-\k_s)$ where $k_l \ll k_s$.

Using the argument above, we can predict the consequences of modifying the
local ansatz to include some form of scale dependence; our discussion here is
similar to that in \cite{Wagner:2010me, Schmidt:2010gw, RomanInProgress}. To get a
feel for the effect of the scale-dependent non-Gaussianity on the bias, notice
that we can rewrite the expression for the $\Delta N$ type non-Gaussian field
(the first line of Eq.~(\ref{eq:twomodels})) in Fourier space as a sum of Gaussian modes $\Phi_G(k)$ and a non-Gaussian piece $\Phi_B(k)$ designed to recover the single-field model bispectrum:
\begin{eqnarray}
\label{eq:FTzeta2}
\Phi({\bf k})&=&\Phi_G({\bf k})+\Phi_B({\bf k})\\\nonumber
\Phi_B({\bf k})&=&\xi_{s}(k)\int\frac{d^3q_1}{(2\pi)^3}
\int\frac{d^3q_2}{(2\pi)^3}\delta_D^3({\bf q}_1+{\bf q}_2+{\bf k})\Phi({\bf q}_1)\Phi({\bf q}_2)
\end{eqnarray}
where we have dropped the $\delta_D^3({\bf k})$ term which is not important for
this discussion. Now we can use Eq.(\ref{eq:FTzeta2}) to consider the effect
of some long-wavelength perturbation $\Delta\Phi(\k_l)$.
Considering $\k\approx-{\bf q}_2\approx\k_s$ and ${\bf q}_1=\k_l$ in that
expression we see that
\begin{eqnarray}
\Delta \Phi_B(\k_s) &=& 2\xi_s(\k_s)\Delta\Phi(\k_l)\Phi(\k_s) \\
\Delta\delta(\k_s) &=& 2\xi_s(\k_s)\Delta\Phi(\k_l)\delta(\k_s)  
\end{eqnarray}
where in the latter equation, we have used the Poisson equation.  We then sum
over all the short wavelengths to get the total boost in peak height, which
then translates into the halo excess and halo bias. Since short wavelength
modes are essentially modes with length scales up to the scale of the object
(there is a window function in the integration), we see that the presence of
the scale-dependent function $\xi_s(k)$ implies that collapsed objects have a
shift in bias with amplitude given by an effective $\fnl$ roughly on the scale
of the object. That is, if $\xi_s(k)$ (which in this simple case is like
$\fnl(k)$ as in Eq.(\ref{eq:runfNLDeltaN})) increases on small scales, smaller
mass objects will have a larger non-Gaussian correction than very massive
objects.

For the curvaton type model, the second line of Eq.~(\ref{eq:twomodels}), the
effect above comes along with an additional $k$-dependence in the bias:
\begin{eqnarray}
\Delta\delta(\k_s) &=& 2\xi_m(\k_l)\xi_m(\k_s)\Delta\Phi(\k_l)\delta(\k_s). 
\end{eqnarray}
In this case, different mass tracers have a non-Gaussian shift in bias,
$\Delta b$, with an amplitude proportional to an ``effective $\fnl$" set by
their mass (a consequence of $\xi_m({\bf k}_s)$) {\it and} a scale dependence
that goes as $k^{-(2-n_f)}$ (a consequence of $\xi_m({\bf k}_l)$).

\subsection{Alternative derivation of scale-dependent effects}
The intuitive procedure above leads to essentially the same result as the
procedure outlined by Grinstein and Wise \cite{Grinstein:1986en} and further
developed by Matarrese et al.\ \cite{Matarrese:1986et}. For a generic
non-Gaussian distribution, Refs.~\cite{Grinstein:1986en,Matarrese:1986et}
found a way to express the two-point function of peaks in terms of a series
expansion in correlation functions. Motivated by the results of
\cite{Dalal:2007cu}, this expression was recently applied to the case of local
non-Gaussianity by Matarrese and Verde \cite{Matarrese:2008nc}. We can use the
same starting point to consider the effects of our generalized local ansatz,
and express the two-point function for halos of mass $M$
($\xi_{h,M}(|\vec{x}_1-\vec{x}_2|)$) in terms of the $n$-point functions of
the density field smoothed on the associated scale $R$
($\xi^{(n)}_R(\vec{x}_1,\dots, \vec{x}_n)$)
\begin{equation}
\xi_{h,M}(|\vec{x}_1-\vec{x}_2|)=\frac{\nu^2}{\sigma(M)^2}
\xi^{(2)}_R(\vec{x}_1,\vec{x}_2)+\frac{\nu^3}{2\sigma(M)^3}
[\xi^{(3)}_R(\vec{x}_1,\vec{x}_1,\vec{x}_2)+\xi^{(3)}_R(\vec{x}_1,\vec{x}_2,\vec{x}_2)]+\dots
\label{eq:halopower}
\end{equation}
where the dots represent higher order terms (both higher correlation functions
and higher powers of the two- and three-point function). The collapse threshold
$\delta_c$ is contained in $\nu\equiv  \delta_c/\sigma(M)$. Now we can use the Fourier transform of the halo auto-correlation,
Eq.~(\ref{eq:halopower}), to compute the bias:
\begin{eqnarray}
\label{eq:MLB}
P_{M,h}(k,z)&=&\frac{\nu^2(z)}{\sigma(M)^2(z)}P_{\delta, R}(k,z)+
\frac{\nu^3(z)}{2\pi^2\sigma(M)^3(z)}P_{\Phi}(k)M_R(k,z)\\[0.2cm]\nonumber
&&\times\xi_{s}(k_p)[\xi_{m}(k_p)]^2\left(\frac{k}{k_p}\right)^{n_f^{(m)}}\int_0^{\infty} dk_1\;k_1^2P_{\Phi}(k_1)M_R(k_1,z)\int_{-1}^{1} d\mu\; M_R(\tilde{k},z)\\\nonumber
&&\times\left[\frac{P_{\Phi}(\tilde{k})}{P_{\Phi}(k)}\left(\frac{k_1\tilde{k}}{k_p^2}\right)^{n_f^{(m)}}
\left(\frac{k}{k_p}\right)^{n_f^{(s)}-n_f^{(m)}}+\left(\frac{\tilde{k}}{k_p}\right)^{n_f^{(s)}}\left(\frac{k_1}{k_p}\right)^{n_f^{(m)}}\right.\\\nonumber
&&\left.+\:\frac{P_{\Phi}(\tilde{k})}{P_{\Phi}(k_1)}\left(\frac{k_1}{k_p}\right)^{n_f^{(s)}}\left(\frac{\tilde{k}}{k_p}\right)^{n_f^{(m)}}\right]\\\nonumber
&=&\frac{\nu^2(z)}{\sigma(M)^2(z)}P_{\delta, R}(k,z)\left[1+
\frac{4\delta_c}{M_R(k,z)}\xi_{s}(k_p)[\xi_{m}(k_p)]^2\left(\frac{k}{k_p}\right)^{n_f^{(m)}}\mathcal{F}_R(k,n_f^{(s)},n_f^{(m)})\right]\\\nonumber
\end{eqnarray}
where $\tilde{k}^2=k^2+k_1^2+2kk_1\mu\;$ and the redshift independent integral is
\begin{eqnarray}
\label{eq:Fofk}
\mathcal{F}_R(k,n_f^{(s)},n_f^{(m)})&=&\frac{1}{8\pi^2\sigma(M)^2}\int_0^{\infty} 
dk_1\;k_1^2P_{\Phi}(k_1)M_R(k_1)\\[0.2cm]\nonumber
&&\times\int_{-1}^{1} d\mu\; M_R(\tilde{k})
\left[\frac{P_{\Phi}(\tilde{k})}{P_{\Phi}(k)}\left(\frac{k_1\tilde{k}}{k_p^2}\right)^{n_f^{(m)}}
\left(\frac{k}{k_p}\right)^{n_f^{(s)}-n_f^{(m)}}\right.\\\nonumber
&&\left.+\:\left(\frac{\tilde{k}}{k_p}\right)^{n_f^{(s)}}\left(\frac{k_1}{k_p}\right)^{n_f^{(m)}}+\:\frac{P_{\Phi}(\tilde{k})}{P_{\Phi}(k_1)}\left(\frac{k_1}{k_p}\right)^{n_f^{(s)}}\left(\frac{\tilde{k}}{k_p}\right)^{n_f^{(m)}}\right]\\\nonumber
&&\rightarrow\frac{1}{2\pi^2\sigma(M)^2}\int_0^{\infty} 
dk_1\;k_1^2P_{\Phi}(k_1)M^2_R(k_1)\left(\frac{k_1}{k_p}\right)^{n_f^{(s)}+n_f^{(m)}}\;.
\end{eqnarray}
The second expression is in the small $k$ limit, so $\tilde{k}\approx
k_1$. In that limit, $\mathcal{F}_R(k,n_f^{(s)},n_f^{(m)})$ is a
constant that depends on the smoothing scale. When $n_f^{(s),(m)}=0$, this
expression reduces to that of \cite{Matarrese:2008nc} and is identically one
in the small $k$ limit.

The Lagrangian halo bias $b_L$ for halos of mass M is defined by
\begin{equation}
\label{eq:biasSquared}
P_h(k)=b_L^2P_{\delta}(k)=b_{L,0}^2\left(1+\frac{\Delta b}{b_{L,0}}\right)^2P_{\delta}
\end{equation}
where in the second equality the $\fnl=0$ contribution $b_{L,0}$ has been
explicitly factored out. Then from Eq.(\ref{eq:MLB}) the change in the bias
relative to the Gaussian value is
\begin{eqnarray}
\label{eq:deltab}
\Delta b&\approx&\frac{\delta_c}{\sigma(M)^2(z)}\left[\frac{2\delta_c}{M_R(k,z)}\xi_{s}(k_p)[\xi_{m}(k_p)]^2\left(\frac{k}{k_p}\right)^{n_f^{(m)}}\mathcal{F}_R(k,n_f^{(s)},n_f^{(m)})\right]\\\nonumber
\end{eqnarray}

\begin{figure}[t]
\begin{center}
\includegraphics[width=0.47\textwidth]{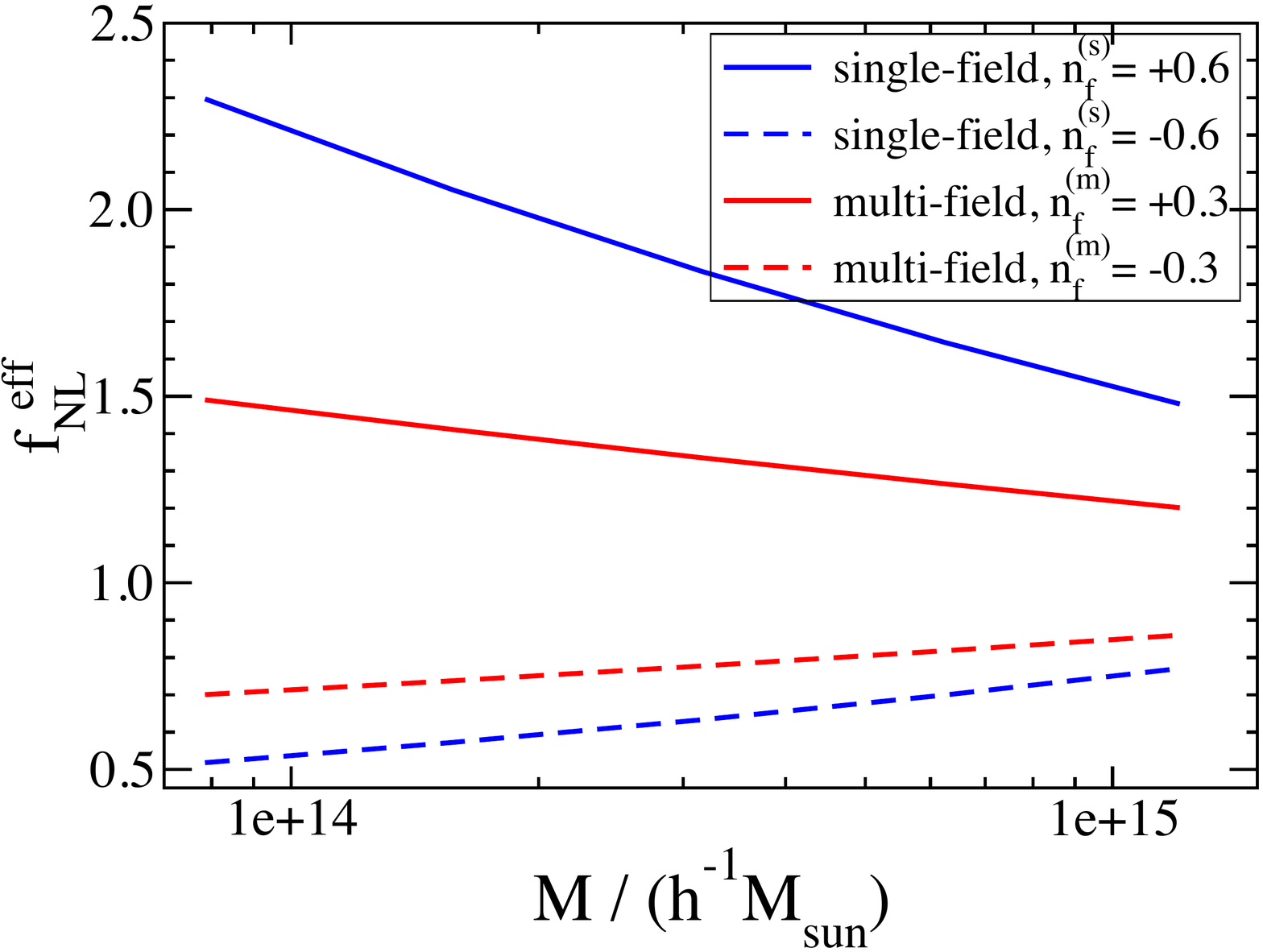} \hspace{0.2cm}
\includegraphics[width=0.47\textwidth]{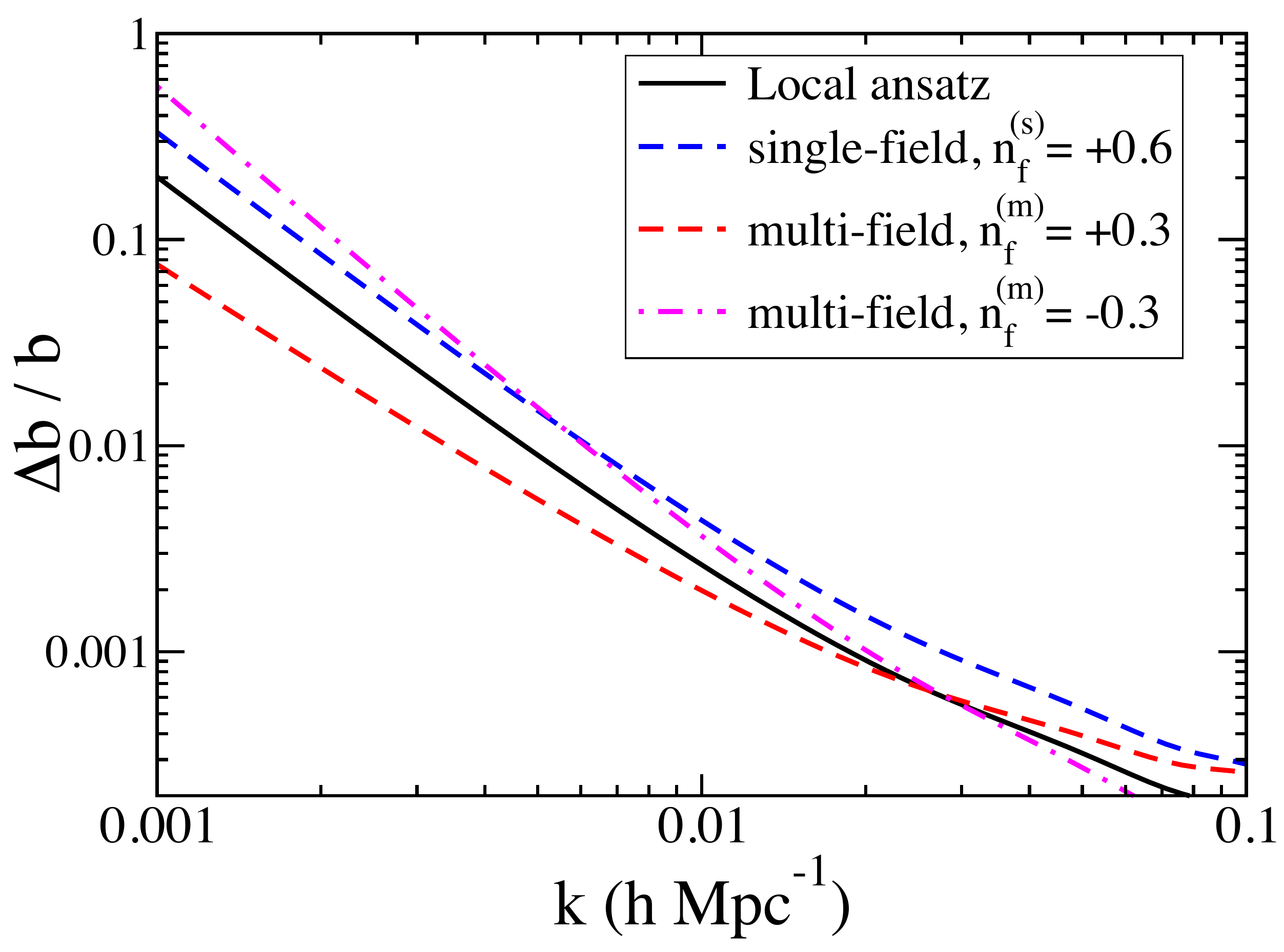} 
\caption{ {\it Left panel}: The effective amplitude of the non-Gaussian bias
  on small scales ($\fnl^{\rm eff}$) as a function of the object's mass for
  two modifications of the local ansatz. The blue short dashed lines are the
  single field model (only $\xi_s(k)$ different from one) and the red long
  dashed lines are the multi-field (only $\xi_m(k)$ different from
  one). The upper lines show the effect of non-Gaussianity that increases on
  small scales, with $n_f^{(s)}=0.6$ or $n_f^{(m)}=0.3$ while the lower lines
  have $n_f^{(s)}=-0.6$ or $n_f^{(m)}=-0.3$. All curves are normalized to
  $\xi_{s,m}(k_p)=1$. {\it Right panel}: A comparison of the correction to the
  bias of objects of mass $4.4\times10^{14}\;h^{-1}\Msun$. The solid black
  curve is the usual local ansatz, the blue long dashed curve is the single-field model with $n_f^{(s)}=0.6$, the red short dashed curve is the multi-field scenario with $n_f^{(m)}=0.3$, and the purple dot-dashed
  curve is the multi-field scenario with $n_f^{(m)}=-0.3$. Again,
  $\xi_{s,m}(k_p)=1$. \label{fig:fNLeff}}
\end{center}
\end{figure}

In using the definition of $\Delta b$ above to derive Eq.~(\ref{eq:deltab}) from Eq.~(\ref{eq:halopower}), we have expanded the square-root which is not always strictly valid. However, the resulting
expression agrees with the peak background split (and we find that keeping the
square-root yields worse agreement between theory and simulation). The results
for the constant, small $k$ part of the integral in Eq.~(\ref{eq:Fofk}),
$\mathcal{F}_R(k)\rightarrow\mathcal{F}(M, k\ll1)$ for the representative
two-parameter cases are plotted as a function of mass (related to smoothing
scale $R$ following Eq.~(\ref{eq:mass_scale})) in Fig.~\ref{fig:fNLeff}. The
functions $\xi_{s,m}$ are normalized to $\xi_{s,m}(k_p)=1$ so that the left
panel shows an effective $f_{NL}$ generated by the scale-dependence for each
scenario. The right panel compares the prediction for the non-Gaussian
correction to the (Lagrangian) bias for the single-field and multi-field scenarios. (This label indicates how many fields contribute to the curvature perturbations - the inflaton itself may be separate).

\subsection{Summary of analytic results}

We have arrived at the same prediction in both of the previous subsections: a
general factorizable and symmetric extension of the local ansatz leads to two
possible modifications of the non-Gaussian halo bias. First, different mass
objects may see a different non-Gaussian correction that goes roughly like the
amplitude of the non-Gaussianity on the scale of the object. Second, the power
of $k$ appearing in the scale-dependent correction can be shifted away from
the standard $k^{-2}$ result when there are two fields contributing to the
curvature power and their relative importance is a function of scale. Either
the first effect alone or a combination of both may be found, depending on the
origin of the scale dependence. The most general case has two parameters to
characterize the running, and one to characterize the amplitude. From the
point of view of measurements of bias, these combine into the mass-dependent
coefficient of the scale-dependent term, $\fnl^{\rm eff}$, and the power of
$k$ that appears in the denominator. In other words, phenomenologically we
have (in the small $k$ limit)
\begin{eqnarray} 
\Delta b_{NG}(k, M)\propto\frac{\fnl^{\rm eff}(M)}{k^{2-n_f^{(m)}}}\;.  
\end{eqnarray}
More precisely (and in terms of the  Gaussian Eulerian bias $b^E_G$)
\begin{equation}
\Delta b_{NG}(k, M)=\fnl^{\rm eff}(M,n_f^{(s)},n_f^{(m)}, k_p)\left(\frac{k}{k_p}\right)^{n_f^{(m)}}\left[
\frac{3(b^E_G-1)\delta_c\Omega_mH_0^2g(\infty)}{c^2k^2T(k)D(z)g(0)}\right]
\label{eq:phenofNL}
\end{equation}
where
\be
\fnl^{\rm eff}(M,n_f^{(s)},n_f^{(m)}, k_p)=\xi_{s}(k_p)[\xi_{m}(k_p)]^2\mathcal{F}_R(k\ll1,n_f^{(s)},n_f^{(m)})\;.
\ee

There is some suggestion, both from simulations and from analytic
considerations, that there is an additional factor multiplying the expression
above for $\Delta b_{\rm NG}$ even in the case of constant local
non-Gaussianity. For example, Giannantonio and Porciani
\cite{Giannantonio:2009ak} have suggested a multiplication by a factor
\begin{equation}
q=1+\frac{\Delta b_I}{b_G^E-1}
\end{equation}
where $\Delta b_I$ is a second order non-Gaussian correction that can be calculated from some choice of non-Gaussian mass function (and the subscript $I$ indicates that it is
scale-independent)\footnote{We thank Tommaso Giannantonio for detailed
  correspondence on this point.}. Although it is reasonably well motivated, we
do not find that such a correction alone substantially improves the fit to our
simulations (especially for negative $\fnl$), so we remain agnostic about the
analytic form of any additional corrections and instead focus on the effects
unique to the generalized local form, especially $\fnl^{\rm eff}(M)$. From a practical perspective, the coefficient above can be fit from simulation and will not affect our conclusions. 

\subsection{Forecasts based on the analytic prediction}

We now estimate the ability of future observations to detect slow-roll values
of the running parameter. Here we present Fisher matrix forecasts based on the
analytic predictions above. This analysis complements earlier forecasts made
for both the scale-independent
\cite{Dalal:2007cu,Carbone:2008iz,Sartoris:2010cr,Cunha:2010zz}, and also
scale-dependent \cite{Sefusatti:2009xu} models of non-Gaussianity.

We first consider the two simpler scenarios, given in Eq.(\ref{eq:twomodels}), that
isolate the single field or multi-field effects and each have only two
parameters. These can be obtained by setting either of the functions $\xi_s$
or $\xi_m$ to one in the general expressions above (and so either $n_f^{(s)}$
or $n_f^{(m)}$ is set to zero). The explicit expressions are 
\begin{eqnarray}
\label{eq:phenofNLexact}
{\rm single\;\: field:}&\:\:\:&\Delta b_{NG}(k, M)=\fnl^{\rm eff}(M,n_f^{(s)}, k_p)\left[\frac{3(b^E_G-1)\delta_c\Omega_mH_0^2g(\infty)}{c^2k^2T(k)D(z)g(0)}\right]\\\nonumber
&&\fnl^{\rm eff}(M, n_f^{(s)},k_p)=\frac{\xi_s(k_p)}{2\pi^2\sigma(M)^2}
\int_0^{\infty} dk_1 \;k_1^2P_{\Phi}(k_1)M^2_R(k_1)\;
\left(\frac{k_1}{k_p}\right)^{n_f^{(s)}}\\\nonumber
{\rm multi-field:}&\:\:\:&\Delta b_{NG}(k, M)=\fnl^{\rm eff}(M,n_f^{(m)}, k_p)\left(\frac{k}{k_p}\right)^{n_f^{(m)}}\left[\frac{3(b^E_G-1)\delta_c\Omega_mH_0^2g(\infty)}{c^2k^2T(k)D(z)g(0)}\right]\\\nonumber
&&\fnl^{\rm eff}(M, n_f^{(m)},k_p)=\frac{\xi_m(k_p)^2}{2\pi^2\sigma(M)^2}
\int_0^{\infty} dk_1 \;k_1^2P_{\Phi}(k_1)M^2_R(k_1)\;
\left(\frac{k_1}{k_p}\right)^{n_f^{(m)}}
\end{eqnarray}
We have used only the small $k$ portion of the integral expression from
Eq.(\ref{eq:Fofk}) since this corresponds to the prediction from the
peak-background split, and since the integrand at high wavenumbers (e.g.\ 
$k\sim\mathcal{O}(0.1)\hmpcinv$) depends on the explicit form of the
window function. We report constraints on the momentum dependent
functions that contribute to the integral in $f_{NL}^{\rm eff}(M)$:
\begin{eqnarray}
\label{eq:fnlk_forms}
{\rm most\;\: general:}&\:\:\:&f_{NL}(k)=\xi_s(k_p)[\xi_m(k_p)]^2\left(\frac{k}{k_p}\right)^{n_f^{(s)}+n_f^{(m)}}\\\nonumber
{\rm single\;\: field\;\: only:}&\:\:\:&f_{NL}(k)=\xi_s(k_p)\left(\frac{k}{k_p}\right)^{n_f^{(s)}}\\\nonumber
{\rm multi-field\;\: only:}&\:\:\:&f_{NL}(k)=[\xi_m(k_p)]^2\left(\frac{k}{k_p}\right)^{n_f^{(m)}}
\end{eqnarray}
where the last two lines specialize to the simpler cases of considering only
the single-field or multi-field effects.  Conceptually, $\fnl^{\rm eff}(M)$
and its Fourier-space analogue, $\fnl(k)$ capture the {\em high
  frequency} scale-dependence of non-Gaussianity, $k\sim M^{-1/3}$.
In addition to this, the bias has low-frequency scale-dependence
$\propto k^{-2+n_f^{(m)}}$ for $k\ll M^{-1/3}$.  This is why our
expressions for $\fnl(k)$ and $\fnl^{\rm eff}(M)$ contain only one
power of $n_f^{(m)}$, even though the bispectrum has two
$k^{n_f^{(m)}}$ terms. Fiducial values adopted
were $\fnl(k_p) \equiv \xi_s(k_p)\xi_m^2(k_p)=30$ and $n_f^{(s),(m)}=0$ while
$k_p=0.04\,{\rm Mpc}^{-1}$ as before.  We will see in a moment that the true
best-measured scale from the large-scale clustering of galaxies and
clusters is somewhat smaller.

Suppose that we have measurements of the power spectrum using objects
(galaxies and clusters of galaxies) that have been separated in several mass
bins.  We assume that the covariance matrix of measured Fourier-space 
overdensities in a given redshift bin centered at $z$ is given by
\begin{equation}
C_{ab}(k, z) = b(k, M_a, z)\, b(k, M_b, z)\,P(k, z) + \delta_{ab}\, \frac{1}{n_a(z)}
\label{eq:cov_massbins}
\end{equation}
where the labels $a$ and $b$ refer to mass bins.  This equation encodes how to
combine observations from different mass bins, and also straightforwardly
specifies the dependence on the parameters of interest $\fnl^{\rm eff}(k_p)$ and $n_f^{(s),(m)}$ via
Eqs.~(\ref{eq:deltab}) and (\ref{eq:Fofk}).

The Fisher matrix can now be evaluated in the FKP approximation \cite{FKP},
where information is summed over the redshift bins and wavenumber shells. We have
\begin{equation}
F_{ij} = \Omega_{\rm survey}
\int_0^{z_{\rm max}}  \left (\frac{dV}{d\Omega dz}\right ) dz\,
\int_{k_{\rm min}}^{k_{\rm max}} 
{\rm Tr} \left [C^{-1}C_{,i}C^{-1}C_{,j}\right ]
\,\frac{k^2 dk}{(2\pi)^2},
\label{eq:Fish_integrals}
\end{equation}
where $\Omega_{\rm survey}$ and $z_{\rm max}$ are the solid angle and maximum
redshift in the survey respectively, $V$ is volume, commas denote
derivatives with respect to the non-Gaussian parameters, and we have
suppressed the dependencies of $C$ on wavenumber and redshift.  In practice we
replace integrals with sums to evaluate this expressions. We neglect the effect of
  redshift uncertainties, but assume thick redshift bins  with $\Delta z=0.2$.

For definiteness, we assume a dataset of the quality expected from the Dark
Energy Survey (DES; \cite{DES}), with $z_{\rm max}=1$ and covering 5000 square
degrees; the total volume in this survey is about $6.5\hinvmpc^3$. We assume
$k_{\rm min}=0.0001\hmpcinv$ and $k_{\rm max}=0.1\hmpcinv$; the latter ensures
that all information safely comes from the linear regime. Finally, we choose
the number density of sources above some mass to correspond exactly to the
expectation from the Jenkins mass function \cite{Jenkins:2000bv}. Therefore,
$n_a(z) = \int_{M_{a, {\rm low}}}^{M_{a, {\rm high}}} (dn/d\ln M)(z)\, d\ln
M$, where ${M_{a, {\rm low}}}$ and $M_{a, {\rm high}}$ and the boundaries of
the $a$-th mass bin. The total number density of sources at $z=0$ and above
$10^{13.5}\Msunhinv$ is $n\simeq 10^{-4}\, (\hmpcinv)^{-3}$. We assume a large number of mass bins (forty) in $M/\Msun$, uniformly distributed
in $\log_{10}M$ from $10^{13.5}\Msunhinv$ to $10^{15.5}\Msunhinv$. 

\begin{figure}[t]
\begin{center}
\includegraphics[width=0.47\textwidth]{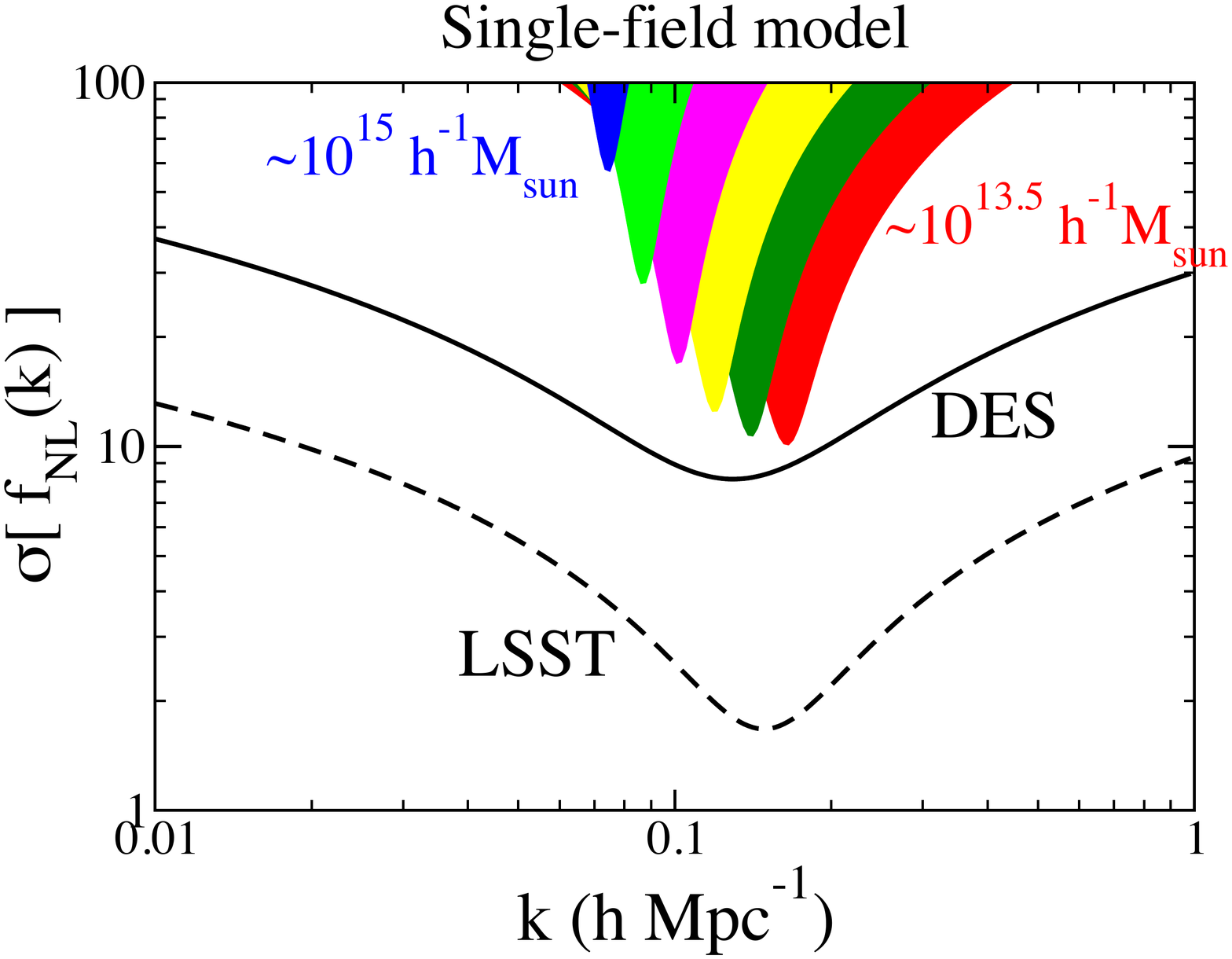}
\includegraphics[width=0.47\textwidth]{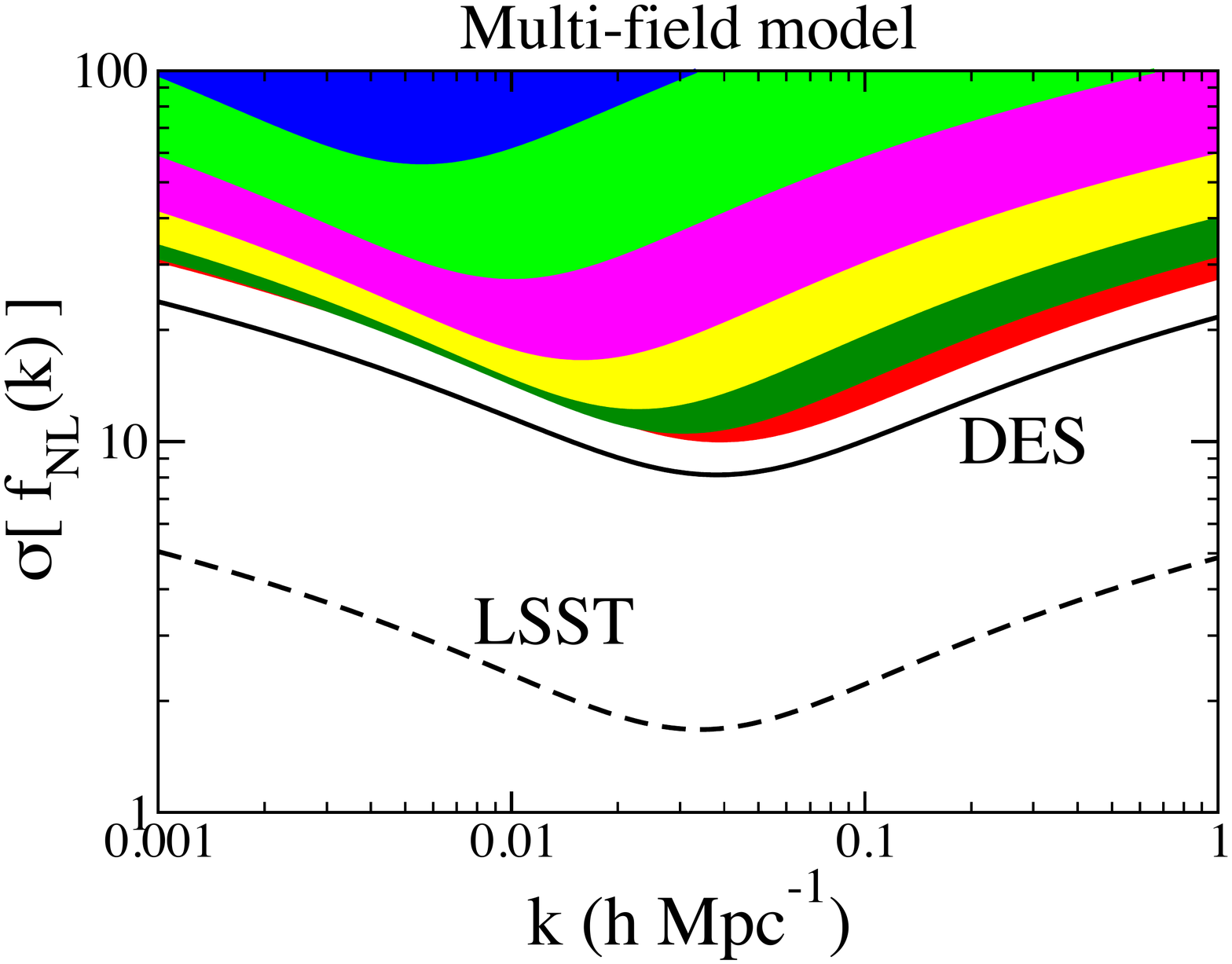}
\end{center}
\caption{Forecasted constraints on $\fnl(k)$ in the model where a single field generates the curvature perturbations ({\it left panel}) and in the multi-field model ({\it right panel}).  We show
  forecasts for data expected from DES (solid black curve) and LSST
  (dashed black curve) observations. The wavenumber at which the
  constraints are the best is $k_{\rm uncorr}$, and at this wavenumber the
  normalization and slope of the power law are precisely uncorrelated.  The
  six colored contours on top of each panel show the individual constraints
  from six narrow mass bins uniformly distributed in $\log_{10}M$ from
  $10^{13.5}\Msunhinv$ to $10^{15}\Msunhinv$ (assuming the DES survey).  In
  the single-field scenario, individual masses do not break degeneracy between
  amplitude and running of $\fnl(k)$ and only constrain this function at a
  single $k$ value; combined masses are required to break the degeneracy. In
  the multi-field scenario, the degeneracy is broken even with halos of a
  fixed mass. [Note that, in all cases, the overall constraints on $\fnl(k)$
    between different wavenumbers $k$ are strongly correlated, given that we
    are assuming a power law in $k$.]  }
\label{fig:Fisher}
\end{figure}

First considering the single-field case, the error in
$\fnl(k)$ at any $k$ is given by a simple propagation of errors
\begin{eqnarray}
\sigma(\fnl(k)) &=& \left [
\frac{\partial \fnl(k)}{\partial \fnl(k_p)} {\rm Cov}_{\rm ff}+
\frac{\partial \fnl(k)}{\partial n_f^{(s)}} {\rm Cov}_{\rm nn} +
2\frac{\partial \fnl(k)}{\partial \fnl(k_p)} 
\frac{\partial \fnl(k)}{\partial n_f^{(s)}} {\rm Cov}_{\rm nf}\right ]^{1/2}\\[0.2cm]
&=& \left [
\left (\frac{k}{k_p}\right )^{n_f^{(s)}}  {\rm Cov}_{\rm ff} +
\fnl(k_p) \left (\frac{k}{k_p}\right )^{n_f^{(s)}} 
 \ln\left (\frac{k}{k_p}\right ) {\rm Cov}_{\rm nn} \right.\\[0.2cm]\nonumber
 &&\left.+
2 \fnl(k_p)\left (\frac{k}{k_p}\right )^{2n_f^{(s)} }\ln\left (\frac{k}{k_p}\right ) {\rm
  Cov}_{\rm nf} \right ]^{1/2}\nonumber
\end{eqnarray}
where ${\rm Cov}\equiv F^{-1}$ is the covariance matrix of the two
non-Gaussian parameters that we consider.  The errors in $\fnl(k)$ are shown
in the left panel of Figure \ref{fig:Fisher}.

We are also interested in finding the best-constrained wavenumber, $k_{\rm
  uncorr}$.  When $k_p=k_{\rm uncorr}$, then the errors on the parameters
$\xi(k_p)$ and $n_f^{(s),(m)}$ are uncorrelated .  While this best-constrained
wavenumber can obviously be read off from Fig.~\ref{fig:Fisher}, it can also
be calculated analytically as
\begin{equation}
k_{\rm uncorr} = k_p \exp 
\left (-\frac{{\rm Cov}_{\rm nf}}{\fnl(k_p){\rm Cov}_{\rm ff}}\right ),
\end{equation}
where $k_p=0.04\hmpcinv$ is the arbitrary pivot in
Eq.~(\ref{eq:runfNL}).  The way that $k_{\rm uncorr}$ `runs' with changing
mass illustrates the point we made in Sec.~\ref{subsec:peak_back} that
different mass halos probe scale-dependent NG on scales corresponding to those
masses.

We find that the best-constrained wavenumber of our survey, for the
single-field model  
and assuming DES-quality data, is $k_{\rm uncorr}\simeq 0.1 \hmpcinv$, and
the corresponding parameter errors at $k_{\rm uncorr}$ are
\begin{equation}
\sigma(\fnl(k_{\rm uncorr})) \simeq  8, \quad
\sigma(n_f^{(s)}) \simeq  0.5 \qquad {\rm (DES\,\, forecast,\,\, single-field)}.
\end{equation}
We also find that the error in $\fnl(k_{\rm uncorr})$ is largely
insensitive to the fiducial value of $\fnl(k_p)$, while the error in
the spectral index $n_f^{(s)}$ becomes larger for a smaller fiducial
$\fnl(k_p)$ (which is expected, since a larger
fiducial non-Gaussianity increases the absolute change in 
$\fnl(k\neq k_{\rm uncorr})$ for a fixed change in $n_f^{(s)}$).
  
We repeated the same exercise assuming data of the quality expected from the
Large Synoptic Survey Telescope (LSST; \cite{LSST}), with $z_{\rm max}=3$ and
covering 20,000 square degrees; the constraints became
\begin{equation}
\sigma(\fnl(k_{\rm uncorr})) \simeq  1.7, \quad
\sigma(n_f^{(s)}) \simeq  0.17 \qquad {\rm (LSST\,\, forecast, \,\, single-field)}.
\end{equation}

Next we consider the multi-field model; see the right panel of
Fig.~\ref{fig:Fisher}. As expected, the numerical constraints on the amplitude
are comparable to the single-field case, however the constraints on
the running improve, and the best-determined scale moves to a slightly
lower $k$: 
\begin{eqnarray}
\sigma(\fnl(k_{\rm uncorr})) \simeq    8, \quad
\sigma(n_f^{(m)}) \simeq  0.2  &\qquad& {\rm (DES\,\, forecast,\,\, multi-field)}\\[0.2cm]
\sigma(\fnl(k_{\rm uncorr})) \simeq    1.7, \quad
\sigma(n_f^{(m)}) \simeq  0.04 &\qquad& {\rm (LSST\,\, forecast,\,\, multi-field)}.
\end{eqnarray}
The colored contours in Fig.~\ref{fig:Fisher} show the individual constraints from each of the six
narrow mass bins uniformly distributed in $\log_{10}M$ from
$10^{13.5}\Msunhinv$ to $10^{15}\Msunhinv$ (this is for the DES survey
scenario and the single-field inflaton model)\footnote{The careful reader will notice that these six bins in
  mass are an oversampling of the 30 original bins in mass we assumed in this
  interval, which are a subset of the total of 40 bins in mass we assumed in
  $M=[10^{13.5},10^{15.5}]\Msunhinv$.}. The thick black curve in either panel
shows the combined constraint. Note that the combined constraint contains the
information from the individual bins {\it and} the correlations between them
(corresponding to $a\neq b$ in Eq.~(\ref{eq:cov_massbins})). A particularly interesting feature of testing these models with primordial
non-Gaussianity is that halos of different mass complement in producing the
overall constraint. For example, inspection of Eqs.~(\ref{eq:deltab}) and
(\ref{eq:Fofk}) shows that, with a single mass measurement, the normalization
and slope of the single-field model, $\fnl(k_p)$ and $n_f^{(s)}$, are
completely degenerate, and only $\fnl(k)$ at a single $k$ value is measured. By
adding a wide range of masses, this degeneracy is broken.
This is shown in the left panel
of Fig.~\ref{fig:Fisher}, where narrow mass bins only constrain this function near a single $k$ value. In contrast, the right panel shows that for the multi-field scenario the degeneracy is broken even with halos of a fixed
mass, as expected from Eq.~(\ref{eq:phenofNL}).

\begin{figure}[t]
\begin{center}
\includegraphics[width=0.47\textwidth]{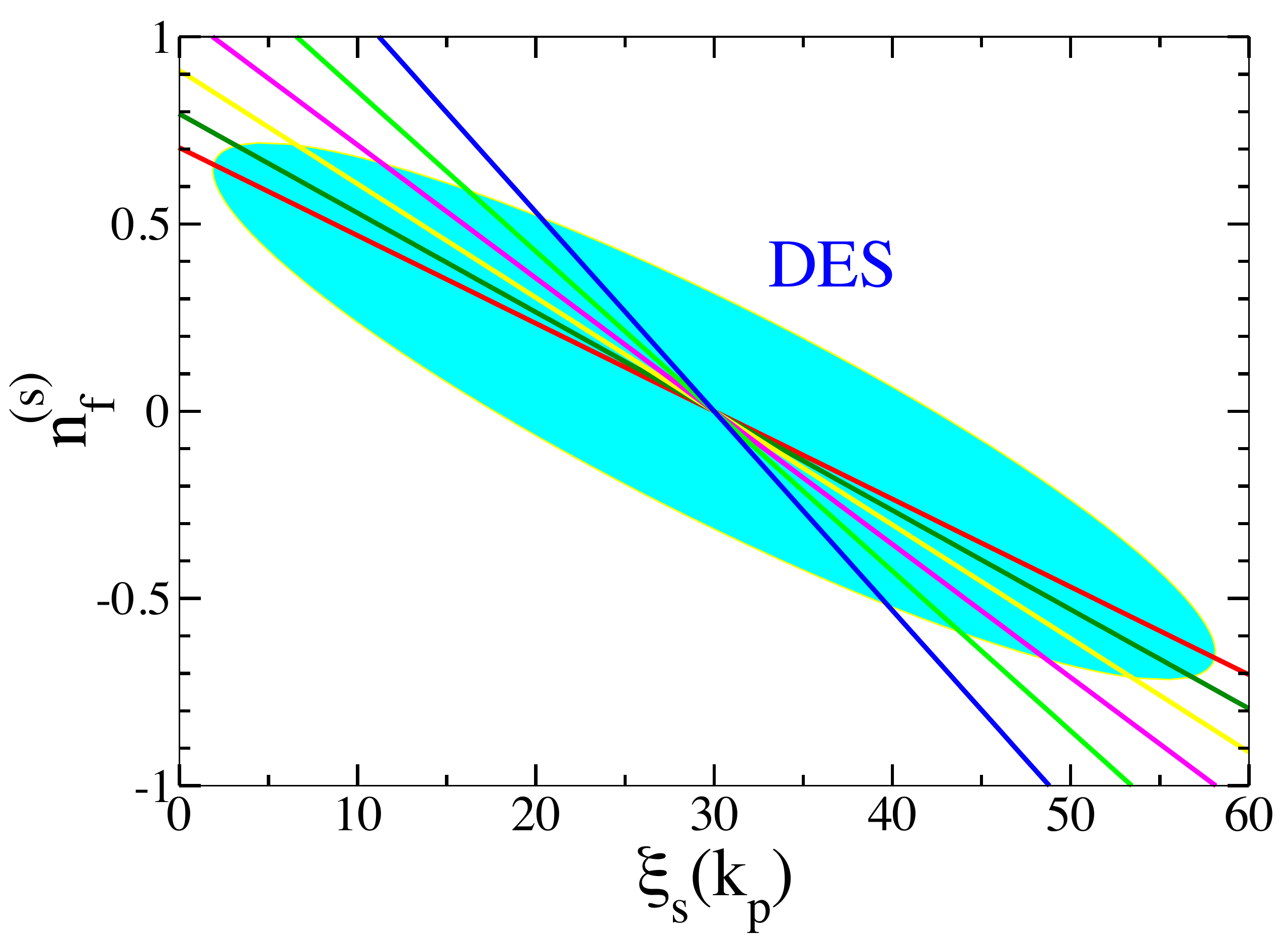} 
\includegraphics[width=0.47\textwidth]{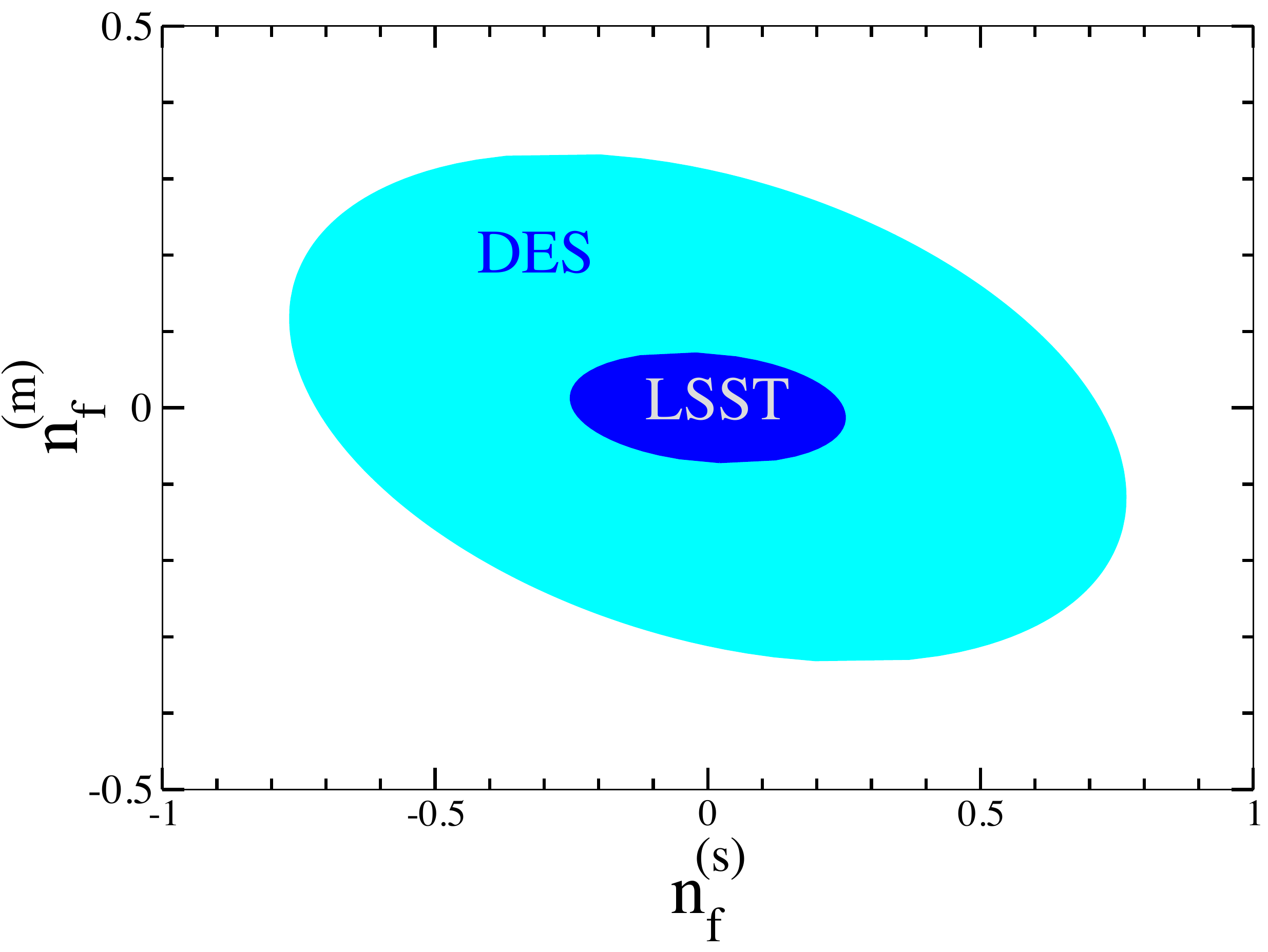} 
\end{center}
\caption{{\it Left panel:} constraints in the $\fnl(k_p)$-$n_f^{(s)}$ plane in the
  inflaton model. Lines show degeneracy directions that each of six individual
  mass bins suffers (these mass bins correspond to colored curves in
  Fig.~\protect\ref{fig:Fisher}). 
{\it Right panel:} Constraints in the $n_f^{(s)}$-$n_f^{(m)}$ plane assuming both
single-field and multi-field models, and marginalizing over the amplitude
(term $\fnl(k_p)\equiv \xi_s(k_p)\xi_m(k_p)^2$ in Eq.~(\protect\ref{eq:fnlk_forms})).  
}
\label{fig:ellipses}
\end{figure}

Similarly, Fig.~\ref{fig:ellipses} contains more visual information on how the
degeneracy is broken with multiple mass measurements. The left panel shows the
constraint in the $\fnl(k_p)$-$n_f^{(s)}$ plane for the DES survey, with lines
showing degeneracy directions that each of the six individual mass bins
suffers (these mass bins correspond to colored curves in
Fig.~\ref{fig:Fisher}). The right panel shows the constraint in the
$n_f^{(s)}$-$n_f^{(m)}$ plane, marginalized over the amplitude
$\fnl(k_p)\equiv \xi_s(k_p)\xi_m(k_p)^2$, for both DES and LSST surveys.

Clearly, even the information from large-scale structure alone offers
the possibility of distinguishing the origin of primordial
non-Gaussianity by constraining both single-field and multi-field model
parameters simultaneously, but the most interesting level to probe is running of order the spectral index,
$n_f^{(s,m)}\sim\mathcal{O}(0.04)$. It is not completely clear if we can reach that level, and there are several factors that could push the predictions above in either direction. First, the forecasts presented here are in some sense a best-case scenario,
given that for simplicity we did not marginalize over the standard
cosmological parameters, and we assumed no systematic errors in recovering the
power spectra of halos, only taking into account the statistical
uncertainties. In particular, measurements of mass of clusters of galaxies
suffer from statistical and systematic errors that are currently at least at
the $10\%$ level per cluster. On the other hand, constraints presented might
be reached in the near future because we do not expect that serious
degeneracies exist between the non-Gaussian and other cosmological parameters
\cite{Dalal:2007cu}.  One exception might be the Gaussian bias $b_G^E$, which
will need to be measured concurrently rather than predicted by theory as we
assumed here. 

In addition, we caution that our simulation
results do not agree quantitatively with the analytic prediction (see the next
section). In fact, the simulations find a substantially {\it stronger} dependence on
mass that what is predicted. If the simulations are proven correct, then the
effects shown above will be easier to distinguish, which is very encouraging
for distinguishing between different mechanisms that may lead to
large local non-Gaussianity.

Finally, the constraints above are from bias alone, but the Planck
satellite will considerably improve constraints on $f_{NL}$ from the CMB
through measurements of the bispectrum itself, so that any scale dependence
may also be constrained \cite{Sefusatti:2009xu}. Whereas current constraints
come from $k\simeq 10^{-3}\hinvmpc$, Planck constraints will extend to
higher $\ell$ and should overlap with constraints from the
bias. Extant CMB analyses \cite{Sefusatti:2009xu} have used a
different parametrization of possible scale-dependence, and so it would be
interesting to repeat this with our ansatz. As 
pointed out by \cite{Tseliakhovich:2010kf} and explored in
detail by \cite{Smith:2010gx}, even more information can be extracted from
joint constraints on models where two fields contribute to the curvature
fluctuation. Multiple observations can separate the inherent size of the
non-Gaussian interaction in one field ($f_{NL}^{\sigma}$) from the rescaling
by the fraction of power from the non-Gaussian field, $\xi(k)$, which combine
as shown in Eq.~(\ref{eq:runfNLcurv}) to give the amplitude of the
non-Gaussian term in the bias, $\xi_m^2(k)=f_{NL}^{\sigma}\xi^2(k)$, that we
have constrained here.

\section{Simulation results}\label{sec:sim}

To check the dependence of the effective $\fnl$ on the tracer mass, we
generated initial conditions with a non-zero bispectrum of the form shown in
the first line of Eq.(\ref{eq:twomodels}) (the single field model) with
scale-dependent amplitude as defined in Eq.~(\ref{eq:runfNL}). This is
equivalent to Eq.~(\ref{eq:runfNLDeltaN}), so our function $\xi_s(k)$
corresponds to a commonly used definition for $f_{NL}(k)$ (see also Eq.~(\ref{eq:fnlk_forms})). This is the
simplest possible model, but serves to check the predictions for how the
effective coefficient of the non-Gaussian bias (``$\fnl^{\rm eff}$") varies
with the mass of the halo. We have also performed a small number of
simulations using the bispectrum form in the second line of
Eq.(\ref{eq:twomodels}), to verify that the bias has the expected
scale-dependence.  These simulations confirm that in such models $\Delta b$ no
longer simply scales as $k^{-2}$ on large scales, but has $n_f^{(m)}$
dependence as well. However, at large enough values of the running to be
distinguished by our simulations, second order effects are significant. For
now we focus on the first scenario which is simpler and already uncovers a
disagreement between the analytic predictions and the numerical results.

To perform these non-Gaussian simulations, we first generated a realization of 
a Gaussian random field $\Phi(x)$ with amplitude chosen so that
$\sigma_8=0.8$ and with spectral index $n_s=0.96$. Then we squared the field,
Fourier transformed and multiplied by the scale-dependent $\fnl$ shown in the second line of
Eq.~(\ref{eq:fnlk_forms}). Finally, we transformed this component back to real
space and added it to the Gaussian piece. We evolved the resulting
non-Gaussian field forward from the scale factor of $a=0.005$ using a
flat cosmology consistent to WMAP7 best fit values ($\Omega_m=0.27$,
$h=0.7$). 

We ran 8 realizations each of Gaussian and non-Gaussian initial conditions,
including $\xi_s(k_p)\equiv\fnl(k_p)=100$ with $n_f^{(s)}=0, 0.6$,
$\fnl(k_p)=300$ with $n_f^{(s)}=0, \pm0.6$, and $\fnl(k_p)=630$ with
$n_f^{(s)}=0, -0.6$. All cases had pivot point $k_p=0.04\; {\rm
  Mpc}^{-1}$. The box size was 2400 $h^{-1}$ Mpc, with $(1024)^3$ particles,
giving a mass per particle of $9.65\times10^{11}h^{-1} M_\odot$. Although our
low-mass halos do not have many particles in these simulations, we used a few
$L_{\rm box}=520 h^{-1}\,$Mpc simulations (with $f_{NL}(k_p)=300$) where these
halos were well-resolved to verify our results. These simulations were
performed on the SciNet machines, where each run took about 3.5 hours on 16
nodes.

We find that the simulations with constant $\fnl$ are offset from the analytic expectation at
small $k$ by a factor that is nearly constant with mass and is less than one
for both positive and negative $\fnl$. This is consistent with findings by other
simulations, and the behavior of the offset was studied in detail by Pillepich
et al.\ \cite{Pillepich:2008ka} and Giannantonio and Porciani \cite{Giannantonio:2009ak}. However, as discussed above, we will effectively fit this offset out and examine only the difference in behavior between our $f_{NL}$ constant simulations and those with running.

\begin{figure}[t]
\begin{center}
\includegraphics[width=0.47\textwidth]{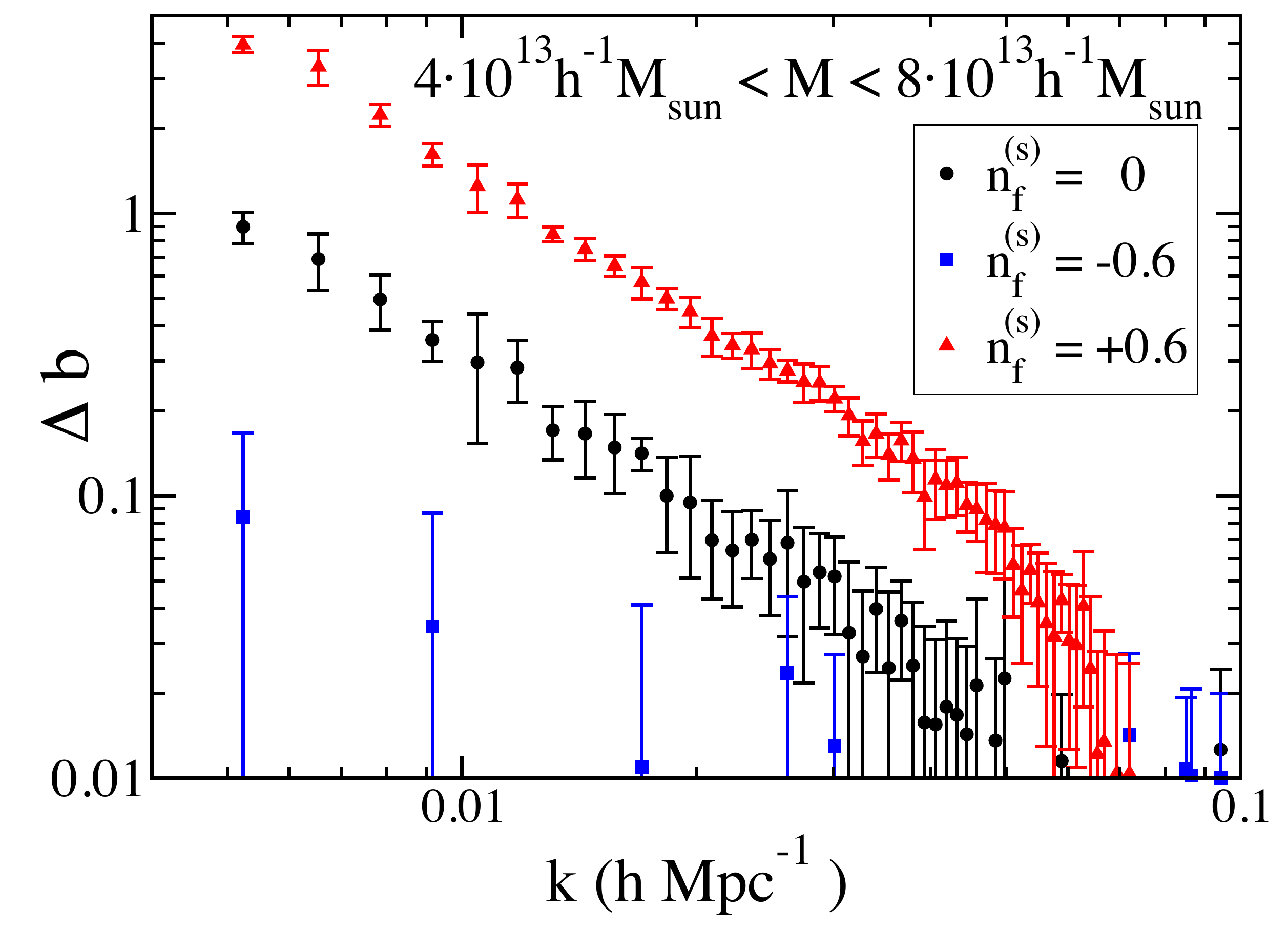}\hspace{+0.2cm} 
\includegraphics[width=0.47\textwidth]{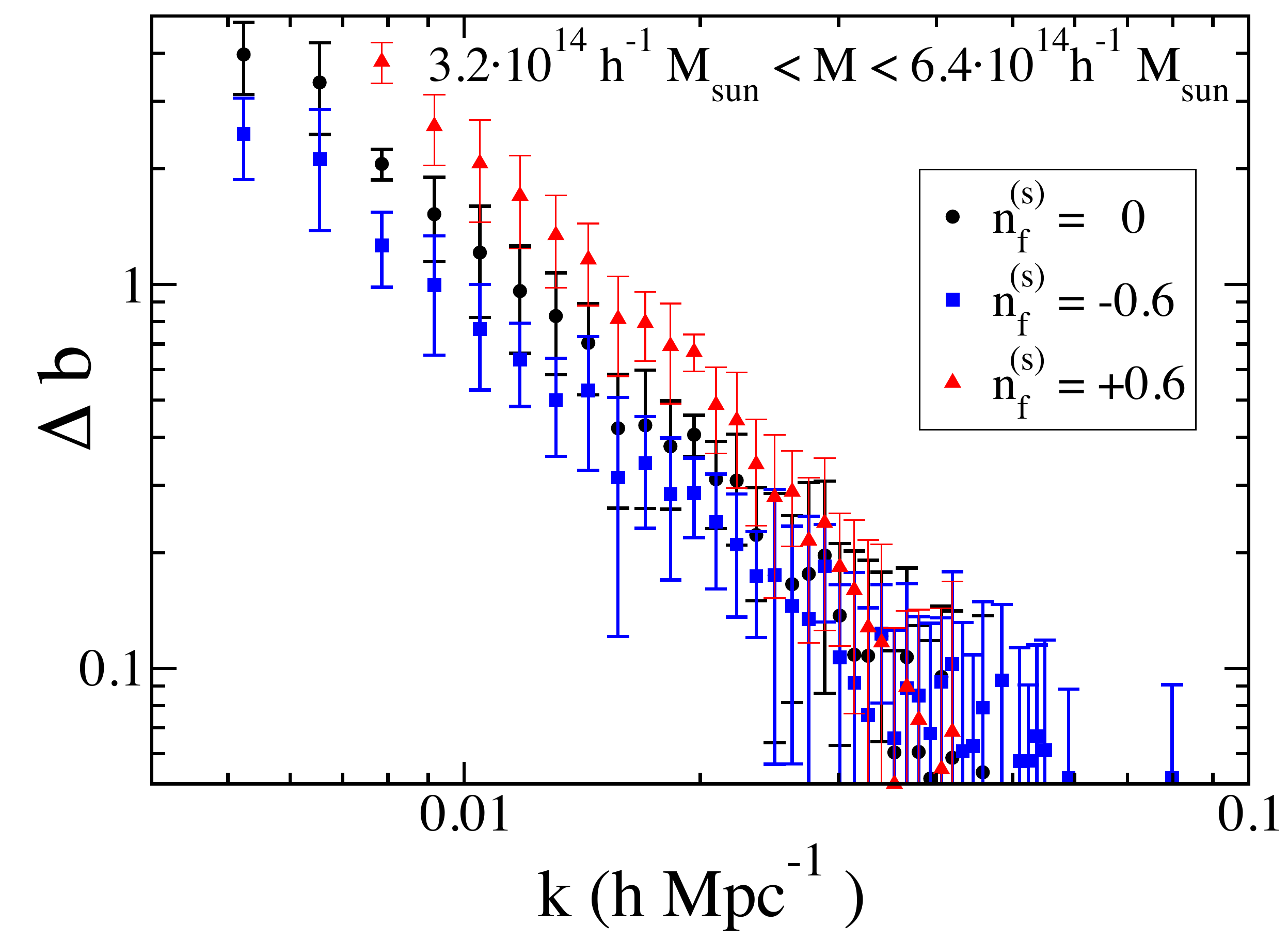} 
\caption{Dependence of scale-dependent non-Gaussian bias on mass, inferred
  from simulations. {\it Left panel}: Simulation results for the non-Gaussian
  contribution to the bias of halos with mass $4-8\times
  10^{13}h^{-1}\Msun$. The black circles points have constant
  $\xi_{s}(k_p)\equiv f_{NL}(k_p)=300$, the blue squares have the same $\xi_s(k_p)$ but
  $n_f^{(s)}=-0.6$, and the red triangles have $n_f^{(s)}=0.6$. Error bars are
  sample variance across several simulations with the same parameters. {\it Right
  panel}: The same set of curves for halos with mass $32-64\times
  10^{13}h^{-1}\Msun$. The scatter here is larger than in the previous plot
  since there are fewer objects at this
  mass. \label{fig:Simresults1} }
\end{center}
\end{figure}

From the simulations with scale-dependent non-Gaussianity, we find that different mass objects are indeed sensitive to an effective
$\fnl$ that depends on the scale of the object and which increases (decreases)
for positive (negative) running as the mass and size of the object decreases.
Figure \ref{fig:Simresults1} illustrates this effect: the non-Gaussian term in
the bias for small mass objects has a smaller (larger) amplitude for positive
(negative) running than for constant $f_{NL}$ (left hand panel). The curves
converge for larger mass objects (right hand panel). The bias correction
$\Delta b$ is calculated from the difference between the matter-halo cross
correlation in a Gaussian simulation and the non-Gaussian case built from the same Gaussian realization, then averaged over realizations.

\begin{figure}[t]
\begin{center}
\includegraphics[width=0.47\textwidth,angle=0]{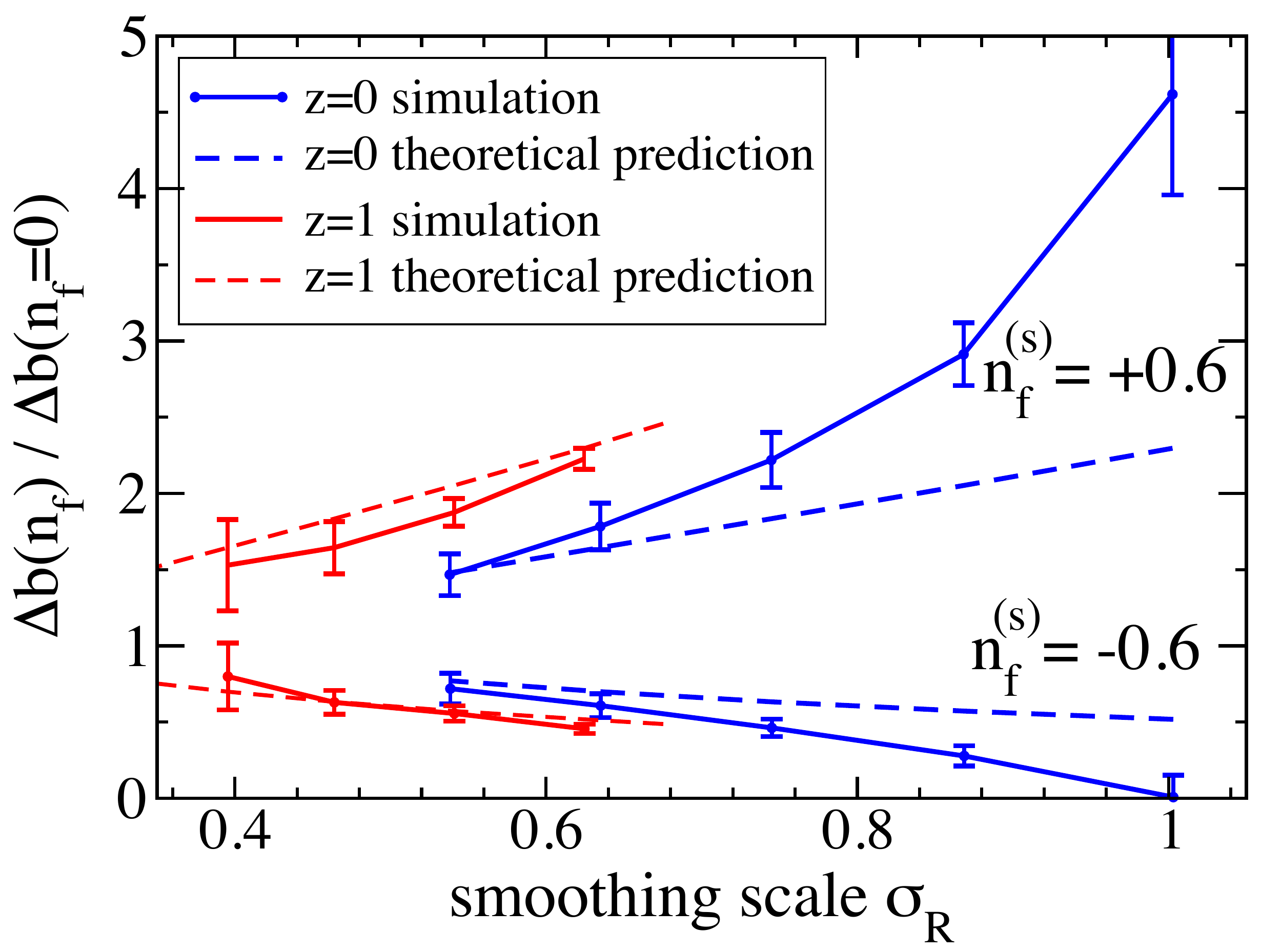} 
\includegraphics[width=0.47\textwidth,angle=0]{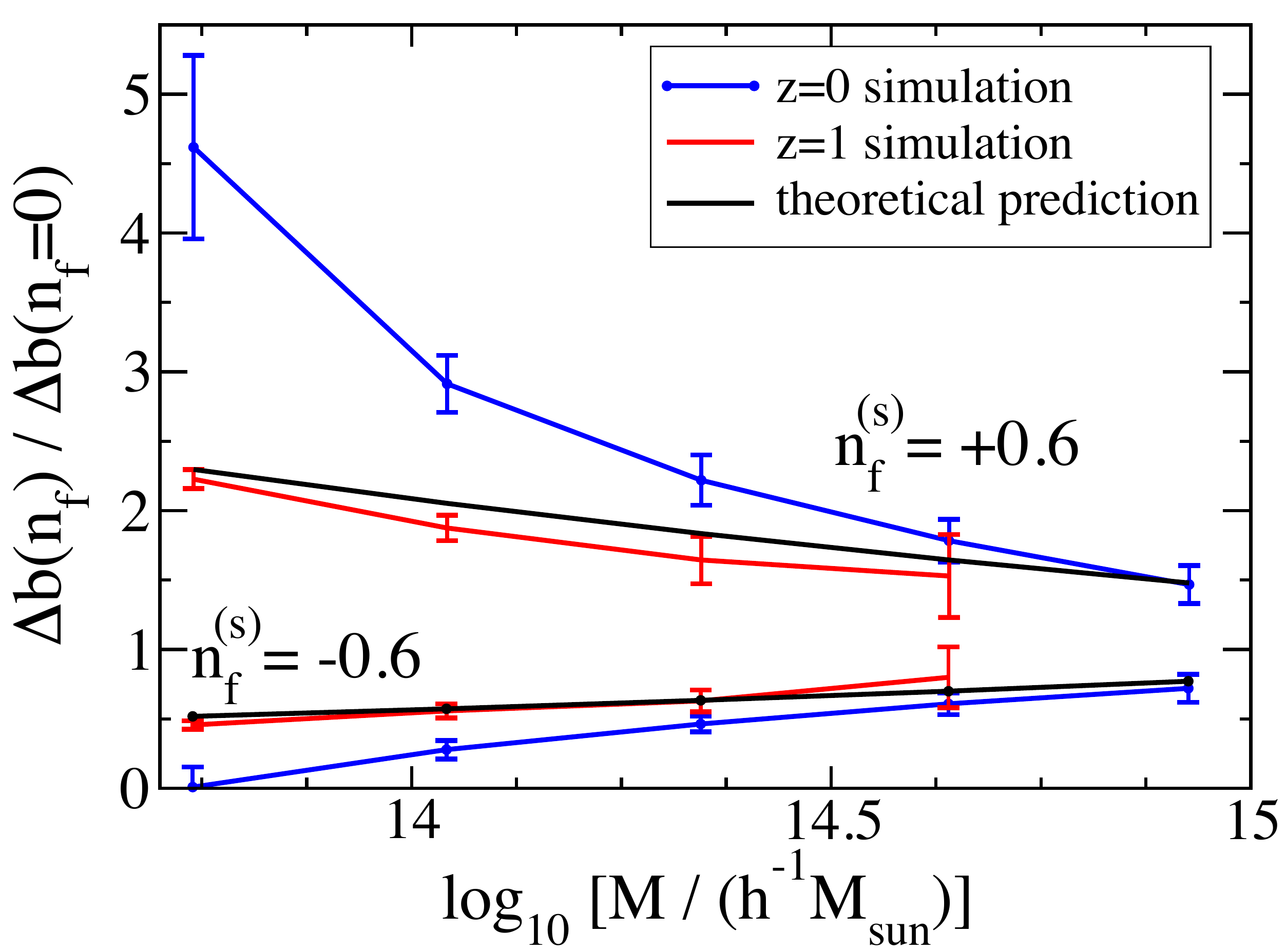} 
\caption{Simulation results for the scale-dependent non-Gaussian bias compared
  to theory. In the {\it left panel}, the vertical axis shows the
  mass-dependent ratio of the bias for non-Gaussianity that runs compared to
  the $f_{NL}$ constant case, measured from $f_{NL}=300$ simulations at
  $z=0$. The upper lines have $n_f^{(s)}=0.6$ and lower lines show
  $n_f^{(s)}=-0.6$. Redshifts $z=0$ (blue, higher values of $\sigma(M)$,and
  $z=1$ (red, lower values of $\sigma(M)$ are shown. The dashed lines are the
  analytical prediction, showing that agreement is better at small
  $\sigma(M)$. The {\it right panel} shows the same information, but plotted
  as a function of mass. Now the theoretical prediction (solid black lines) is
  redshift independent. \label{fig:fnlEffectives}}
\end{center}
\end{figure}

The qualitative effect we expected is present, but for some halos the
magnitude of the effect is not well predicted by the analytic expressions from
Section 3. Figure \ref{fig:fnlEffectives} shows the deviation between
simulation and prediction for $f_{NL}=300$, $n_f^{(s)}=\pm0.6$ (the same
generic trend was seen in all parameter sets). We plot the ratio of the
non-Gaussian correction with running to the non-Gaussian correction for
constant $f_{NL}$: \be \mathcal{F}^{\rm sim}\equiv\frac{b(f_{NL}=300,
  n_f^{(s)}=0.6)-b(f_{NL}=0)}{b(f_{NL}=300,
  n_f^{(s)}=0)-b(f_{NL}=0)}=\frac{\Delta b(n_f^{(s)})}{\Delta
  b(n_f^{(s)}=0)}\;.  \ee This is compared with the theoretical expectation
calculated from the small $k$ limit of Eq.(\ref{eq:Fofk}). The curves are
plotted as a function of $\sigma(M)$. As the figure demonstrates, the
simulation results agree well with our analytic model in the high-mass limit
$\sigma(M) \ll \delta_c$, but towards lower masses (e.g.\ $\sigma(M) \gtrsim
0.8$) the simulations produce a stronger effect than Eq.(\ref{eq:Fofk}) would
predict.  Note that the discrepancy does not appear at a fixed mass, but
rather at a fixed $\sigma(M)$.  This is illustrated in the right panel of
Figure \ref{fig:fnlEffectives}, which is identical to the left panel of Figure
except that now mass $M$ is the abscissa. The figure shows that for fixed mass
$M$, the simulations agree with Eq.(\ref{eq:Fofk}) at high redshift, but begin
to disagree at low redshift as $\sigma(M)$ grows.

One very plausible explanation for this discrepancy at low mass
($\sigma \gtrsim 0.8$) is that the profiles of the peaks that produce
halos begin to change as $\sigma$ increases.  As we have argued, the
non-Gaussian bias of halos of mass $M$ is sensitive to the value of
$f_{NL}$ at some effective $k\propto M^{-1/3}$.  Implicit in this
scaling is the assumption that the profiles of peaks that collapse
into halos are similar at different masses, just rescaled in size.
However, we know that this assumption is incorrect.
Bardeen et al.\ \cite{Bardeen:1985tr} argued from Gaussian statistics
that as $\sigma(M)$ increases, the peaks that collapse into halos
generally become steeper.  N-body simulations confirm the presence of
this effect, but show that it is much stronger in magnitude than
predicted by Bardeen et al., apparently due to environmental effects
during halo formation \cite{Dalal:2010hy}. Because peaks at high $\sigma(M)$
are much steeper than rare peaks at low $\sigma(M)$, they are sensitive
to non-Gaussianity at higher wavenumbers, even at the same peak size
$R$.  For scale-independent $f_{NL}$, this change in peak profile has
no effect, but for nonzero $n_f$ it can dramatically enhance the mass
dependence of non-Gaussianity, as our simulations show.  It remains to
be seen whether the magnitude of the change in peak profile can
account for the discrepancy between our simulations and
Eq.(\ref{eq:Fofk}); this is work in progress.
  
\section{Conclusions}\label{sec:concl}

In summary, we have introduced a generalization of the local ansatz, Eq.~(\ref{eq:SymFact}), that is a symmetric, factorizable function of the momenta and includes scale-dependent
non-Gaussianity. This more general expression is motivated by natural features
of models that give an observably large amplitude for local type
non-Gaussianity, and distinguishes between non-Gaussian curvature fluctuations generated by
a single field and multiple fields. If only one field contributes to the curvature fluctuation (and is different from the inflaton so that the non-Gaussianity may be large), the scale-dependence of the non-Gaussianity characterizes the self-interactions of the field. If two fields contribute to the curvature fluctuations, scale-dependence indicates how the ratio of power in the fields changes, which is a function of how different the potentials are. If local non-Gaussianity is large enough to be observed, such scale-dependence is as natural as running of the power spectrum.

Models with scale-dependent local non-Gaussianity can generate two signatures in the non-Gaussian
contribution to the halo bias. First, the non-Gaussian term may be
proportional to an effective $\fnl$ related to the amplitude of the bispectrum
on the scale of the object so that different mass objects have a different amplitude correction. Second, the $k^{-2}$ behavior of the non-Gaussian
bias can be modified to $k^{-(2-n^{(m)}_f)}$ (where $|n_f^{(m)}|<1$), and one
should expect the first effect to accompany this one.

We have used N-body simulations to verify that different mass objects do
indeed have a non-Gaussian bias proportional to an effective $\fnl$ that
varies with the mass of the object. It is interesting that the simulations show that
scale dependance with $n_f^{(s),(m)}<0$ can erase the scale dependent effect
on the bias for some range of masses, highlighting the need for analysis using multiple tracers of
different mass. However, the quantitative result for halos at large $\sigma(M)$ is not
well predicted by our analytic expressions. We have speculated that the origin of this discrepancy may be related to differences in the initial peak profiles of the halos, but leave a detailed investigation for a later work.

Future surveys are sure to bring interesting results. Using the
analytic predictions, we find that they may be able to distinguish the
different pieces of our generalized local ansatz, and so different
origins of local non-Gaussianity, especially if the running is
somewhat large ($n_f^{(s),(m)}\sim\mathcal{O}(0.1)$). However, the
existing analytic expressions predict a {\it weaker} effect than we
see in the simulations, and our forecasts only account for constraints
from massive groups and clusters of galaxies, neglecting the
(potentially) greater sensitivity to running possible when galaxy
correlations are included as well.  Our forecasts for future surveys
should therefore be taken as a lower limit on the potential to
observationally distinguish these features.

{\bf Acknowledgments} We thank Niayesh Afshordi, Carlos Cunha,
Adrienne Erickcek, Louis Leblond, Fabian Schmidt, Roman Scoccimarro
and especially Chris Byrnes for many very useful discussions and
comments. We are grateful to Olivier Dor\'e for collaboration in the
early stages of this project, and for a careful reading of the manuscript. 
We also thank the Benasque Center for Physics for providing an
excellent work environment for the last stages of this project, as well as the
opportunity for discussions with Vincent Desjacques, Tommaso Giannantonio,
Cristiano Porciani, and Emiliano Sefusatti. Computations were performed on the
General Purpose Cluster at the SciNet HPC Consortium. SciNet is funded by the
Canada Foundation for Innovation under the auspices of Compute Canada; the
Government of Ontario; Ontario Research Fund - Research Excellence; and the
University of Toronto. Research at the Perimeter Institute is
supported in part by the Government of Canada through Industry
Canada and by the Province of Ontario through the Ministry of
Research and Information (MRI). DH is supported by the DOE OJI grant under contract
DE-FG02-95ER40899, NSF under contract AST-0807564, and NASA under contract
NNX09AC89G.

\bibliographystyle{JHEP}

\providecommand{\href}[2]{#2}\begingroup\raggedright\endgroup

\end{document}